%% file: main.tex
\documentclass[a4paper]{article}

\usepackage[utf8]{inputenc}
\usepackage[english]{babel}
\usepackage{csquotes}
\usepackage[backend=biber, style=apa, url=false, doi=false, backend=biber]{biblatex}
\usepackage[textsize=small]{todonotes}
\usepackage{caption}
\usepackage{subcaption}
\usepackage{amsmath,amssymb,amsfonts,bm}
\usepackage{nicematrix}
\usepackage{nameref} % For referencing sections by name

\DeclareMathOperator*{\argmin}{argmin}

\author{Andrey Zhdanov$^{1,4}$, Jussi Nurminen$^2$, Joonas Iivanainen$^3$, Samu Taulu$^{4,5}$\footnote{Corresponding author}}

\title{A Minimum Assumption Approach to MEG Sensor Array Design}

\bibliography{bibliography}

\begin{document}
\maketitle
\noindent
1. BioMag Laboratory, HUS Diagnostic Center, Helsinki University Hospital and University of Helsinki, Helsinki, Finland\\
2. Motion Analysis Laboratory, Children’s Hospital, University of Helsinki and Helsinki University Hospital, Helsinki, Finland \\
3. Sandia National Laboratories, Albuquerque, NM 87185, USA \\
4. Department of Physics, University of Washington, Seattle, WA, USA \\
5. Institute for Learning and Brain Sciences, University of Washington, Seattle, WA, USA
%%-------------------------------------------------------------------------
%% The main body
%%
\begin{abstract}
\textbf{Objective}:
Our objective is to formulate the problem of the Magnetoencephalographic (MEG) sensor
array design as a well-posed engineering problem of accurately measuring the
neuronal magnetic fields. This is in contrast to the traditional approach that
formulates the sensor array design problem in terms of neurobiological interpretability
the sensor array measurements.

\textbf{Approach}:
We use the Vector Spherical Harmonics (VSH) formalism to
define a figure-of-merit for an MEG sensor array. We start with an observation
that, under certain reasonable assumptions, any array of $m$ perfectly noiseless
sensors will attain exactly the same performance, regardless of the sensors'
locations and orientations (with the exception of a negligible set of singularly
bad sensor configurations). We proceed to the conclusion that under
the aforementioned assumptions, the only difference between different array
configurations is the effect of (sensor) noise on their performance. We then
propose a figure-of-merit that quantifies, with a single number, how much the
sensor array in question amplifies the sensor noise.

\textbf{Main results}:
We derive a formula for intuitively meaningful, yet mathematically rigorous
figure-of-merit that summarizes how desirable a particular sensor array design is.
We demonstrate that this figure-of-merit is well-behaved enough to be used as a cost
function for a general-purpose nonlinear optimization methods such as simulated
annealing. We also show that sensor array configurations obtained by such
optimizations exhibit properties that are typically expected of \enquote{high-quality}
MEG sensor arrays, e.g. high channel information capacity.

\textbf{Significance}:
Our work paves the way toward designing better MEG sensor arrays by isolating 
the engineering problem of measuring the neuromagnetic fields out of the bigger
problem of studying brain function through neuromagnetic measurements.

\end{abstract}

\input{introduction.tex}

\input{methods.tex}

\input{results.tex}

\input{discussion.tex}

\input{acknowledgements.tex}

\printbibliography

\end{document}

\maketitle

%%-------------------------------------------------------------------------
%% The main body
%%

\begin{figure}[h]
    \centering
    \includegraphics[width=\textwidth]{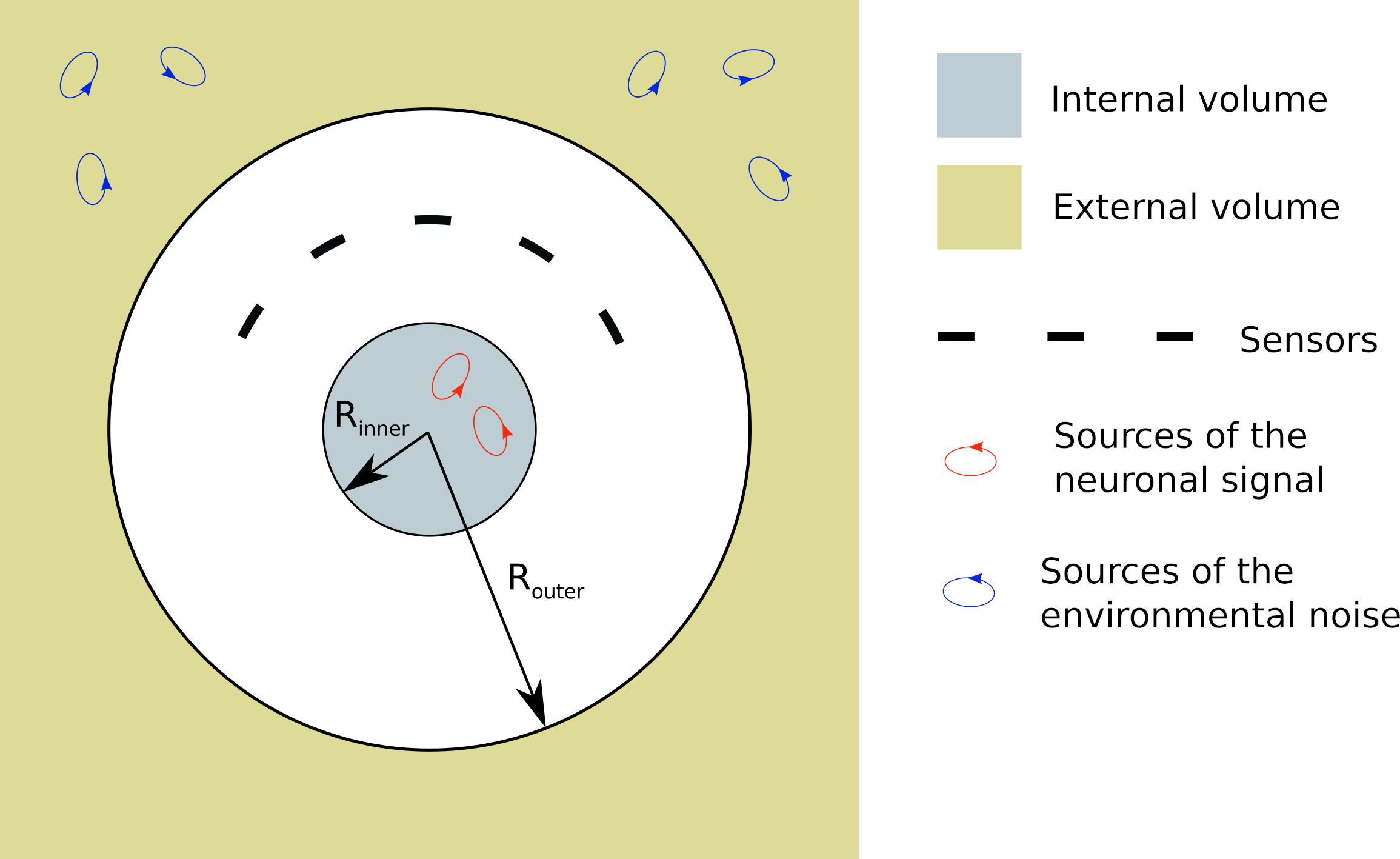}
    \caption{Single-origin VSH model}
    \label{fig:vsh_basic}
\end{figure}

We make the following assumptions: all the currents of interest (brain currents)
are localized within a sphere (the sphere models the head, so we call it the
head sphere) surrounded by a current-free volume that includes the sampling
volume as a subset. Additionally there are some distant current sources
(environmental noise) located much further away from the center of the sphere
than any of the sensors (see Figure~\ref{fig:vsh_basic}). More formally, there
exists a single origin $\mathbf{r_\text{o}}$ and two radii $R_{\text{inner}} < R_{\text{outer}}$,
such that:
\begin{itemize}
	\item all the current sources of interest (the sources whose fields we want
	to measure) lie within the distance $R_{\text{inner}}$ from $\mathbf{r_\text{o}}$, in other
	words, within the volume $\{\mathbf{r}: | \mathbf{r} -\mathbf{r_\text{o}}| < R_{\text{inner}} \}$
	(we call it \emph{inner volume})
	\item all the noise sources lie further than $R_{\text{outer}}$ from the origin,
	in other words, within the volume $\{\mathbf{r}: | \mathbf{r} -\mathbf{r_\text{o}}| > R_{\text{outer}} \}$
	(we call it \emph{outer volume})
	\item there are no current sources within the shell between $R_{\text{inner}}$ and
	$R_{\text{outer}}$, in other words, within the volume
	$\{\mathbf{r}: R_{\text{inner}} < |\mathbf{r}-\mathbf{r_\text{o}}| < R_{\text{outer}¸}\}$ (we call it
	\emph{the source-free volume}), and the sampling volume $V_{\text{samp}}$ is located totally
	within this shell.
\end{itemize} 

For the aforementioned setup, the magnetic field $\mathbf{B}(\mathbf{r})$
everywhere throughout the 3D space is the sum of the magnetic fields
generated by the currents in the inner and the outer volumes:
\begin{equation}
	\mathbf{B}(\mathbf{r}) = \mathbf{B}_\alpha(\mathbf{r}) +
	\mathbf{B}_\beta(\mathbf{r}) \qquad \forall \mathbf{r} \in \mathbb{R}^3,
\end{equation}
where $\mathbf{B}_\alpha(\mathbf{r})$ is the neuronal magnetic field that
we want to measure and $\mathbf{B}_\beta(\mathbf{r})$ is the environmental
noise.

Considering the fact that the neural electromagnetic phenomena can be
accurately modeled by the quasistatic approximation of Maxwell's equations
\parencite{Hamalainen1993}, the magnetic field $\mathbf{B}(\mathbf{r})$ throughout the
source-free sampling volume $V_{\text{samp}}$ is curl-free
\begin{equation}
	\nabla \times \mathbf{B}(\mathbf{r}) = 0
\end{equation}
where $\mathbf{r} \in V_{\text{samp}}$ is the vector specifying
location.\footnote{From here onwards we assume the applicability domain of all the equations to be
the sampling volume $\mathbf{r} \in V_{\text{samp}}$ unless explicitely stated
otherwise.} It has been demonstrated \parencite{Jackson1998, Arfken2013} that
under the aforementioned conditions the magnetic fields $\mathbf{B}_\alpha(\mathbf{r})$
and $\mathbf{B}_\beta(\mathbf{r})$ can be expanded in terms of spherical
harmonic basis functions:
\begin{equation}\label{eq:field_exp}
\begin{aligned}
	\mathbf{B}_\alpha(\mathbf{r}) &= \sum_{l=1}^\infty\sum_{m=-l}^l \alpha_{lm}\mathbf{B}_{\alpha_{lm}}(\mathbf{r})\\
	\mathbf{B}_\beta(\mathbf{r}) &= \sum_{l=1}^\infty\sum_{m=-l}^l \beta_{lm}\mathbf{B}_{\beta_{lm}}(\mathbf{r}),
	\end{aligned}
\end{equation}
where $\alpha_{lm}$ and $\beta_{lm}$ are expansion coefficients, and
$\mathbf{B}_{\alpha_{lm}}$ and $\mathbf{B}_{\beta_{lm}}$
are predefined basis functions that have the following form:

\begin{equation}
\begin{aligned}
	\mathbf{B}_{\alpha_{lm}} &= -\mu_0 r^{-(l+2)}\bm{\nu}_{lm}(\theta, \varphi) \\
	\mathbf{B}_{\beta_{lm}} &= -\mu_0 r^{l-1}\bm{\omega}_{lm}(\theta, \varphi),
\end{aligned}
\end{equation}
where $(r, \theta, \varphi)$, is the representation of $\mathbf{r}$ in spherical
coordinates centered at the center of the brain sphere, $\mu_0$ is the magnetic
permeability of vacuum, and $\bm{\nu}_{lm}(\theta, \varphi)$ and
$\bm{\omega}_{lm}(\theta, \varphi)$ are the vector spherical harmonic
functions\footnote{For the precise definitions of $\bm{\nu}_{lm}(\theta, \varphi)$
and $\bm{\omega}_{lm}(\theta, \varphi)$ see \cite{Taulu2005}.}. Note that
$\bm{\nu}_{lm}(\theta, \varphi)$ and $\bm{\omega}_{lm}(\theta, \varphi)$
do not depend on $r$. The set $\{\mathbf{B}_{\alpha_{lm}} : l=0\ldots\infty, m=-l\ldots l\}$
is called the \emph{internal} basis as it spans the space of all the magnetic
fields originating form the currents in the inner volume. Similarly, the
\emph{external} basis $\{\mathbf{B}_{\beta_{lm}} : l=0\ldots\infty, m=-l\ldots l\}$
spans the space of all the magnetic fields caused by the currents in the outer volume.

As the case with most infinite series expansions, the utility of
Equation~\ref{eq:field_exp} comes from the fact that in real-world scenarios the
magnitudes of $\alpha_{lm}$ and $\beta_{lm}$ decrease pretty quickly with $l$. That is, in practice,
we can restrict ourselves to a relatively small subset of all the basis functions
corresponding to small values of $l$ without introducing any significant errors
into the field estimate:
\begin{equation}
\begin{aligned}
	\mathbf{B}_\alpha(\mathbf{r}) 
	&\approx \sum_{l=1}^{L_\alpha}\sum_{m=-l}^l \alpha_{lm}\mathbf{B}_{\alpha_{lm}}(\mathbf{r})\\
	\mathbf{B}_\beta(\mathbf{r}) 
	&\approx \sum_{l=1}^{L_\beta}\sum_{m=-l}^l \beta_{lm}\mathbf{B}_{\beta_{lm}}(\mathbf{r})
\end{aligned}
\end{equation}
for some relatively small integers $L_\alpha$ and $L_\beta$.

The basis functions $\mathbf{B}_{\alpha_{lm}}(\mathbf{r})$ and
$\mathbf{B}_{\beta_{lm}}(\mathbf{r})$ are perfectly known
and do not depend on the intracranial source distributions or sensor array
configuration. Hence if we know the values of a finite set of coefficients
$\{\alpha_{lm}: 1\leq l\leq L_\alpha, -l\leq m\leq l\} \cup \{\beta_{lm}: 1\leq l\leq L_\beta, -l\leq m\leq l\}$, we
know everything about the magnetic field in the sampling volume.

\printbibliography

%% file: introduction.tex
\section{Introduction}
Magnetoencephalography (MEG) is a noninvasive brain imaging modality that
studies neuronal activity through measurement, outside of the head, of magnetic
fields created by neuronal currents \parencite{Hamalainen1993, Cohen2003}.
Electric currents in the brain (intracranial currents), in accordance with
Maxwell's equations, produce magnetic fields that extend to the volume outside
the head, where they can be measured noninvasively. In MEG, one measures the magnetic
fields with sensors located outside the head and tries to infer
the intracranial currents generating these measurements. The fact that the magnetic
fields outside the head (extracranial magnetic fields) are related to intracranial
currents through Maxwell's equations makes it possible to infer spatiotemporal
features, with limited certainty, about the currents from the signals measured
with MEG sensors.

Ideally, we would like to measure the extracranial fields and compute the
intracranial currents that produced them. Such computation is called
\emph{the inverse problem}. However, in the general case, the inverse problem is
\emph{ill-posed} -- it cannot be solved uniquely. This is because some
intracranial currents produce zero extracranial magnetic field.\footnote{This
statement needs a bit of explanation. An important property of the quasistatic Maxwell
equations is their linearity -- the magnetic fields are related to the intensity
of the currents that cause them by a linear transformation. Any linear
transformation has this property: it is possible to uniquely identify the
transformation's input given its output (i.e. compute the inverse transformation)
if there exists no non-zero input that causes the transformation to produce
zero output. That is why the impossibility to uniquely estimate the intracranial
currents from the extracranial fields is the same thing as existence of a
non-zero intracranial current that produces zero extracranial magnetic field.}
One particularly celebrated example of "a silent current" -- radial current dipole in a
spherically symmetric conductor -- is described in \cite{Sarvas1987}. Obviously,
no extracranial MEG measurement can reveal anything about the silent currents.

Whereas no MEG sensor array (collection of magnetic field sensors) located
outside the head can reveal everything about the intracranial currents, some
sensor arrays can reveal much more than others, depending on the number, locations,
orientations of sensors and other parameters \parencite{Kemppainen1989, Hamalainen1993}.
%\todo{more citations here?}. 

There is another complication to the problem of MEG measurement, a more practical one. So far we
assumed that the extracranial magnetic fields are caused by the intracranial
currents only. However, in any practical situation the measurements of magnetic
fields around the head will be contaminated by environmental noise (magnetic
fields produced by various artificial sources: power grid, elevators, electric
motors operating nearby, etc.). The environmental noise can be, to a considerable
degree, reduced by various shielding techniques (e.g., \cite{taulu2019novel}), however
in any practical MEG setup the residual noise is still non-negligible. Thus the
magnetic fields measured by the sensors are a sum of two components: (a) the
neuronal component -- the magnetic fields produced by the intracranial currents and associated volume currents in the head,
and (b) the environmental
noise produced by much stronger currents located far away from the sensors. We
would like our sensor array to reveal as much as
possible about the the intracranial currents not only in the noiseless case, but
also in the presence of the environmental noise.
% \todo{Citations here, active/passive shielding, full-body superconductive shielding etc.}

All of the above makes MEG sensor array design an important problem that has attracted considerable
attention. One very tempting approach to MEG sensor array design is to try to summarize the
\enquote{goodness} of the array using a single scalar -- figure-of-merit.
Once we find a figure-of-merit that describes well enough the array's ability to
characterize intracranial currents (preferably, in the presence of environmental noise),
sensor array design becomes a multidimensional
non-linear optimization problem -- a problem that has been widely studied, and
for which multiple practical tools are available. Unsurprisingly, a variety of
figures-of-merit have been proposed to date. These include measures such as precision in locating
cortical current sources \parencite{Hamalainen1993,beltrachini2021optimal}, and information about the sources conveyed by the array
\parencite{Kemppainen1989,iivanainen2021spatial, schneiderman2014information}.
%, and others\todo{more citation here}.
%\todo{we need an earlier citation here}

As we mentioned before, the ultimate, albeit unreachable (in the general case) goal
of MEG is to solve the inverse problem. Therefore, it is not surprising that some of
the figures-of-merit proposed to date: (a) make some assumptions about all
possible intracranial currents that improve the conditioning of the inverse problem,
and (b) summarize in a single number the sensor array's ability to solve the inverse
problem under these assumptions. The problem with this approach is that it critically
depends on the accuracy of the assumptions, but there is no good way to ensure such
accuracy. Additionally, previously proposed figures-of-merit generally focus on the
sensor array's performance in the absence of environmental noise
\parencite{Kemppainen1989, Hamalainen1993}. 
%\todo{We definitely need to do a more extensive literature research here!}

In the current paper we propose a novel figure-of-merit for MEG sensor array design that
is not centered around solving the inverse problem. We do not
try to solve an ill-posed problem of characterizing the intracranial currents
through additional assumptions that improve the conditioning. Instead, following the approach by \cite{ahonen1993,grover2016information,iivanainen2021spatial}, we solve
the much less ambitious, but well-conditioned problem of measuring the magnetic
fields outside of the head as accurately as possible. This approach might seem
counterintuitive as it explicitly ignores the inverse problem and instead focuses
on measuring as accurately as possible something that might be of no interest (per se)
to MEG users -- the magnetic field outside the head. Nonetheless, we argue that
separating the question \enquote{What can we say about intracranial currents from
extracranial magnetic field measurements?} from the question \enquote{How can we
measure extracranial magnetic fields as accurately as possible?} makes a lot of
sense from the sensor array designer's perspective.\footnote{Strictly speaking, we
are cheating here a little bit. When we talk about \enquote{\ldots measuring \ldots magnetic fields as
accurately as possible} we implicitly assume that there exists some way to
define what \enquote{accurate} means -- in other words, that there is a 
way to measure the (dis-)similarity between the two magnetic fields (the true one and the
estimate) as a single scalar. In reality there are many different ways to quantify
the difference between two magnetic fields, each leading to a different definition
for figure-of-merit of the sensor array. In our simulations we used the $L^\infty$-norm
to define the difference, but other other choices are equally possible.}
%In reality, magnetic fields being vector functions
%over $\mathbb{R}^3$, there is no single \enquote{true} way of quantifying the
%similarity between two magnetic fields with a single scalar. One can argue that
%different choices for magnetic field similarity measures essentially correspond
%to different implicit assumptions about the statistical properties of the
%intracranial currents. We discuss this furter in the Subsection~\ref{magnetic_field_similarity}.}
The former question necessarily
requires some assumptions about the intracranial currents, which are particularly
problematic during the array design stage since these assumptions are specific to
a particular MEG experiment. The latter question, on the other hand, is independent
of such assumptions and constitutes exactly the question that MEG sensor array
designer should address. %\todo{Should we move a part of this into the discussion?}

As an additional benefit, our approach provides a straightforward and principled way
of incorporating the resilience to the environmental noise
into the figure-of-merit. The question "How can we measure extracranial magnetic 
fields as accurately as possible?" can be naturally extended to the question "How
can we measure the neuronal component of the extracranial fields as accurately as
possible?" without the need for arbitrary weight factors balancing the accuracy of
the inverse problem solution against the noise resilience.
%\todo{is this a good word?}

Briefly stated, our approach consists of using the vector spherical harmonics (VSH)
decomposition \parencite{hill1954theory,Taulu2005} of the magnetic field to define a field model which we use to
optimize the sensor array. Using the VSH decomposition we define cutoff values for
the spherical harmonics degrees $l$ of the inner and outer expansions corresponding
to fields due to neural sources and external interference, respectively. By using
the VSH field model, we investigate how measurement noise maps into magnetic field
interpolation noise for a given sensor array configuration. We define a figure of
merit that quantifies how much the noise gets amplified in the process. We design
sensor arrays that minimize the figure of merit, i.e, that aim not to amplify noise.

%% file: methods.tex
\section{Methods}
\subsection{Array Geometry}
\subsubsection*{Array Geometry Constraints}
When designing an MEG sensor array, we cannot place the sensors completely
freely. For example, we cannot place them inside the head, or too far away
from the head, or too close to each other, etc. We denote the set of all
admissible sensor configurations as $\mathbf{\Xi}$. Each point $\xi\in\mathbf{\Xi}$
is a possible sensor array; $\mathbf{\Xi}$ is the domain of the sensor array
optimization problem.

For the purpose of this paper, we assume point-like sensors that measure
magnetic field along a certain direction (sensor orientation). There are no
constraints on sensor orientations; the only constraint on sensor locations is
that all the sensors are located within a closed volume adjacent to the head,
called \emph{sampling volume} $V_{\text{samp}}$.
Thus $\mathbf{\Xi}$ is uniquely defined by $V_{\text{samp}}$ and the number
of sensors $m$. 
\begin{equation} \label{eq:domain_definition}
    \mathbf{\Xi} \triangleq \{(\mathbf{r}, \mathbf{e}) | \mathbf{r} \in V_\text{samp}, \|\mathbf{e}\|=1\}^m
\end{equation}
where $\mathbf{r}$ and $\mathbf{e}$ denote the location and orientation of a
sensor, respectively. Whereas these assumptions are not perfectly realistic, the
resulting simulations provide important insights into the real-world MEG sensor
array design as we will see in Section~\ref{results}.

\begin{figure}[ht]
    \centering
    \begin{tabular}{cc}
    \includegraphics[width=0.4\textwidth]{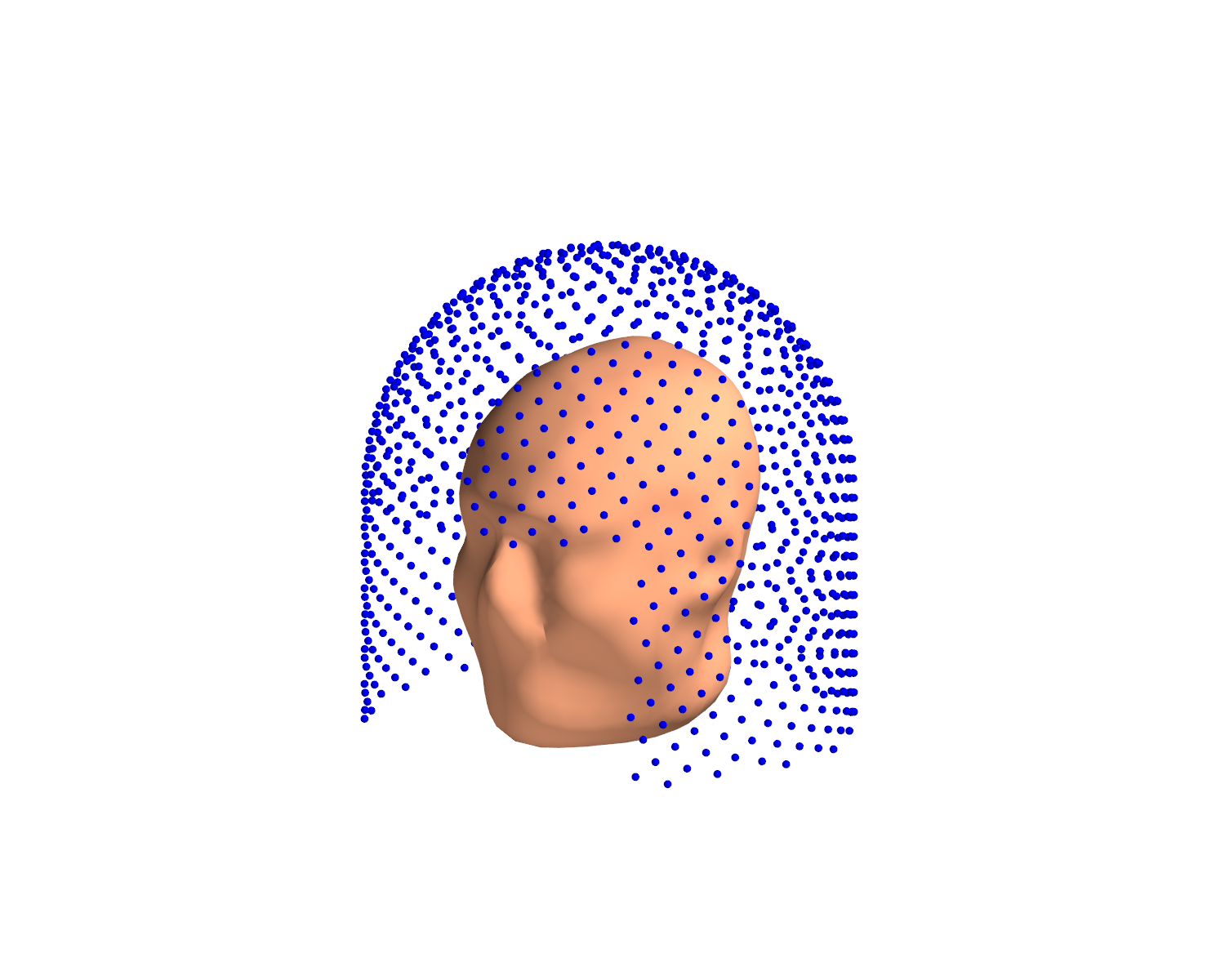} &
    \includegraphics[width=0.4\textwidth]{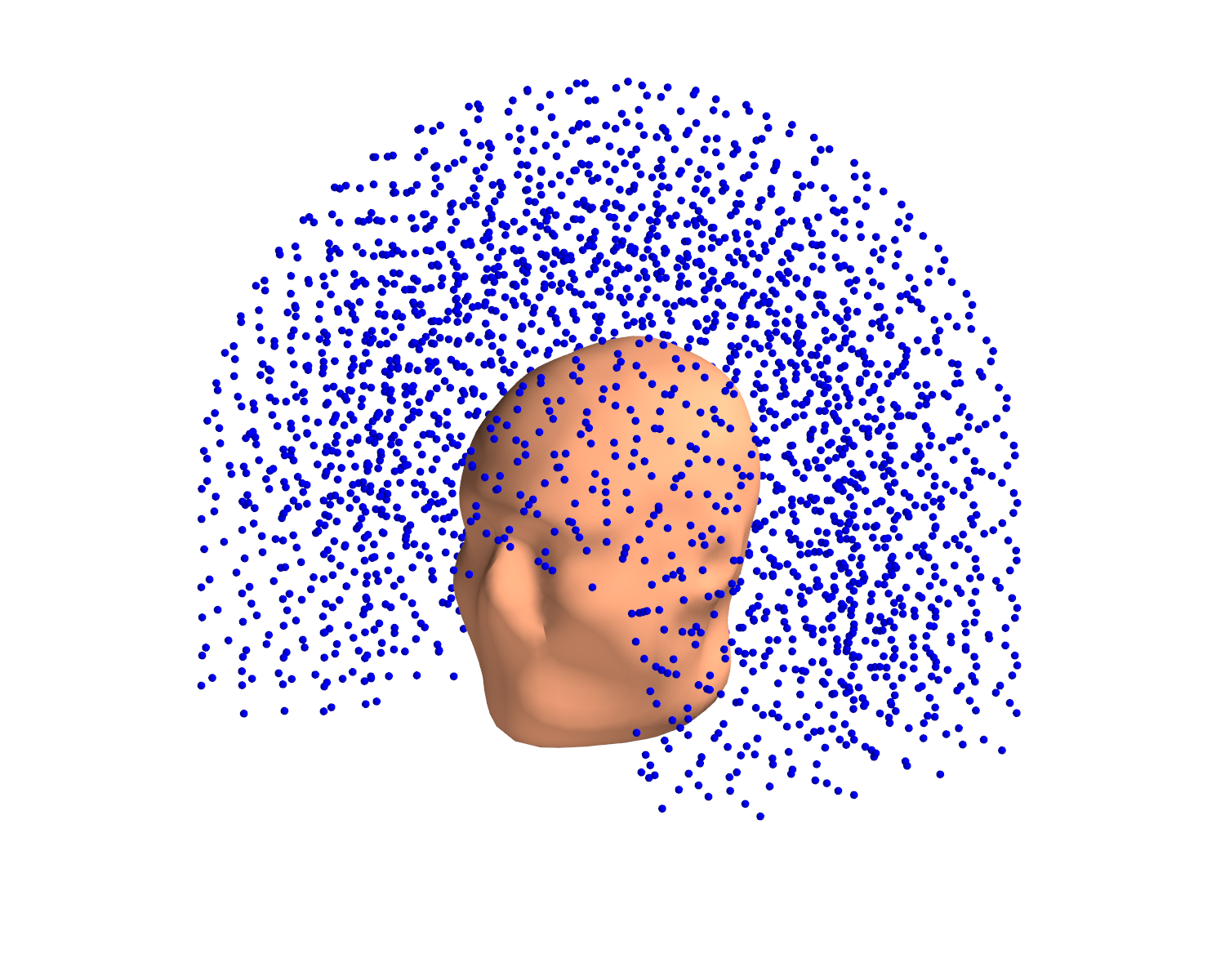}
    \end{tabular}
    \caption{Sampling volume used in our paper. Blue dots are sampling
    locations used for the discretization of the continuous sampling
    volume $\{\mathbf{r}|\mathbf{r}\in V_\text{samp}\}$. for 2D (left)
    and 3D (right) sampling volumes.}
    \label{fig:sampl_vol}
\end{figure}

In this paper we mostly consider two different sampling volumes: a 3D and a 2D.
Both are helmet-shaped, defined as a union of two
geometric primitives (see Figure~\ref{fig:sampl_vol}):
\begin{enumerate}
	\item a section of a cylindrical shell (wrapped around the subject's head
	with an	opening in front of the face), and
	\item a hemispherical shell covering the top of the head
\end{enumerate}

The height of the cylindrical shell was 15 cm, and the opening spanned an
angle of $\pi / 2$ radian. For the 3D volume both primitives have a
finite thickness of 0.1 m
(inner radius of 0.15 m and outer radius of 0.25 m), and for the 2D
volume both have zero thickness with the inner and outer radii being
equal to some value $R$ (and to each other). Thus for the 2D sampling
volume our sensor array is a function of $R$.\footnote{Note that $R$
affects only the radii of the two primitives comprising the sampling
volume, whereas the height of the cylindrical shell is fixed to 0.15 m
independently of $R$. Thus arrays for different values of $R$ are not
scaled versions of each other.}

\begin{figure}[ht]
    \centering
    \begin{tabular}{cc}
    \includegraphics[width=0.4\textwidth]{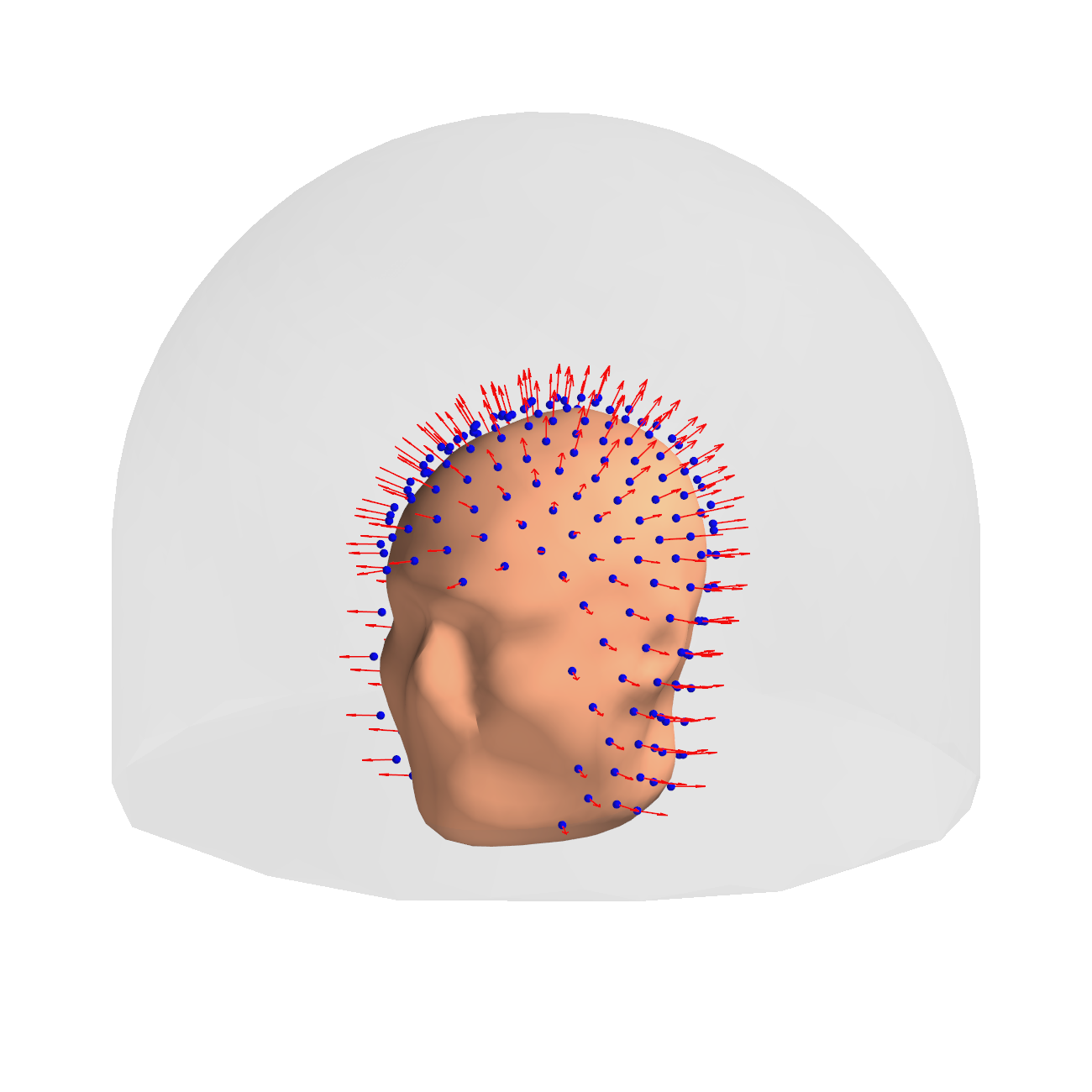} &
    \includegraphics[width=0.4\textwidth]{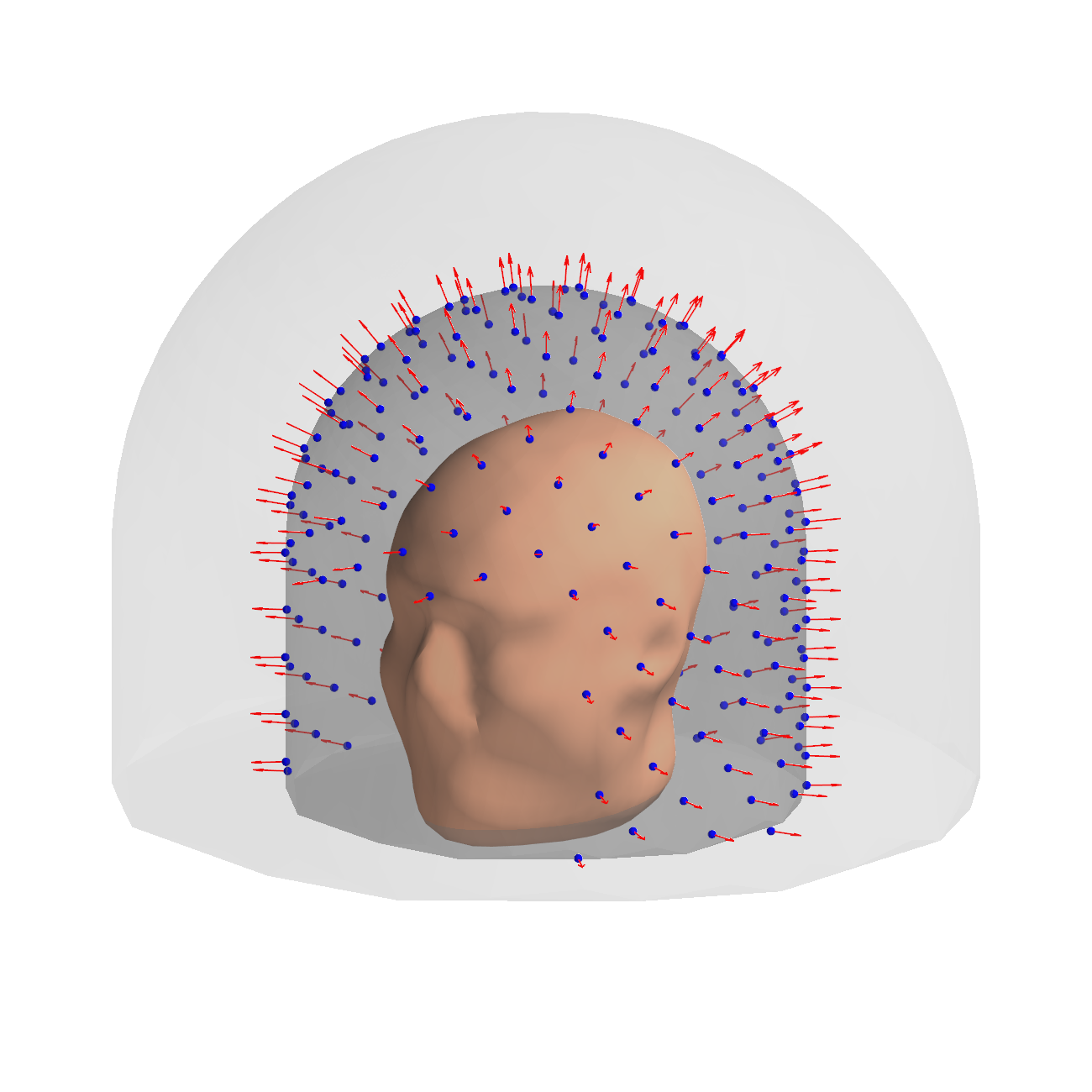}
    \end{tabular}
    \caption{The difference between the anatomically-constrained (left) 
    and regular (right) 3D sampling volumes. In both cases the sensors
    are located on the innermost wall of the sampling volume. Blue spheres mark
    sensor locations; red arrows denote the direction along which the
    magnetic field is measured.}
    \label{fig:scalp_array_example}
\end{figure}

In addition to the 3D sampling volume described above, in some of our
experiments we also use an anatomically-constrained variation of the 3D
sampling volume. The main difference between the two is that in the
anatomically-constrained version the inner wall of the sampling volume
is defined by the subject's individual anatomy -- sensors can be placed
anywhere down to the distance of 7 mm from the MRI-based anatomical
head surface (see Figure~\ref{fig:scalp_array_example}). The offset of
7 mm is based on a typical dimension of an
OPM sensor \parencite{Shah2018}; for the purpose of our simulations we
used an anatomical head surface provided by the MNE-Python package
\parencite{Gramfort2013}.

Note that whereas 3D sampling volume is a closed region of 3D space
of non-zero volume, the 2D volume is a zero-volume surface. Nevertheless,
we will use the term \enquote{volume} for both of them for convenience.

\subsubsection*{Uniformly-Spaced Radial Arrays}
\label{sec:uniformely_spaced_arrays}
We define a special type of MEG sensor array -- an (approximately)
uniformly-spaced radial array -- that we are going to use as an example
of what a reasonable non-optimized MEG sensor array might look like.

\begin{figure}[h]
    \centering
    \includegraphics[width=0.3\textwidth]{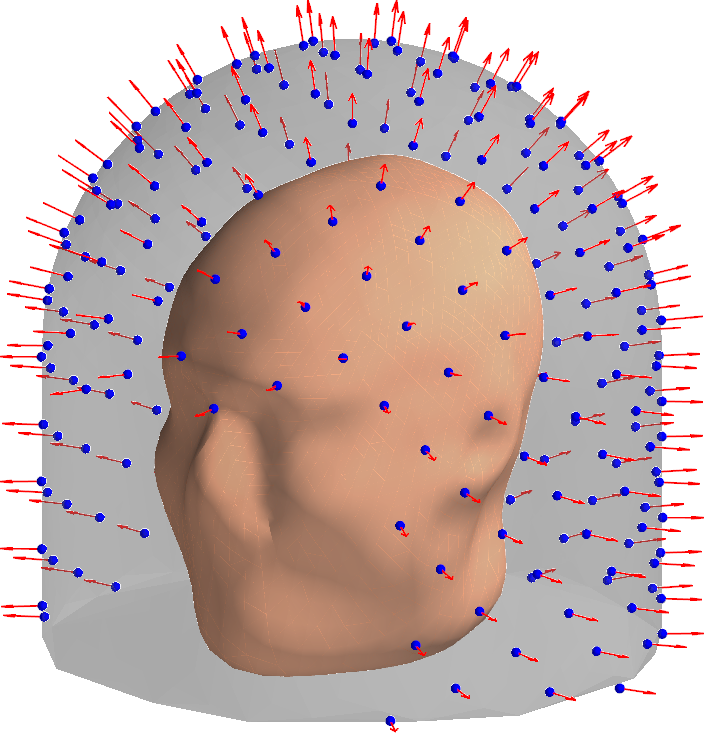}
    \caption{A uniformly-spaced radial sensor array. Blue spheres mark
    sensor locations; red arrows denote the direction along which the
    magnetic field is measured.}
    \label{fig:regular_array}
\end{figure}

The uniformly-spaced radial array of $N$ sensors (see figure~\ref{fig:regular_array})
comprises $N$ sensors
approximately uniformly distributed over the 2D sampling volume of radius $R$.
The points are distributed over the sampling volume (2D surface in this case)
using an algorithm based on the idea of the generalized
spiral set on a sphere \parencite{Saff1997}. The orientations of the
sensors (e.g. the directions along which the magnetic field is measured)
are normal to the sampling volume. Thus uniformly-spaced radial array
is a function of $N$ and $R$.

\subsection{Array Figure-of-Merit Definition}
Let us assume that we are given a sampling volume $V_\text{samp}$ and number of
sensors $m$, thus defining the domain $\mathbf{\Xi}$ of our sensor array
optimization problem. For each particular sensor array configuration
$\xi\in\mathbf{\Xi}$ we have $m$ measurements of the magnetic field at $m$
(possibly distinct) locations within $V_\text{samp}$. At each location
$\mathbf{r}$ we measure a single component of the magnetic field vector
$\mathbf{B}(\mathbf{r})$, along the particular sensor's orientation. We
assume that everywhere throughout $V_\text{samp}$ the value of $\mathbf{B}(\mathbf{r})$
is accurately enough approximated by the first $n$ VSH components
(where $n=L_\alpha(L_\alpha+2) + L_\beta(L_\beta+2)$ for some appropriately chosen
positive integers $L_\alpha$ and $L_\beta$):\footnote{For more details on this,
see the supplementary material.}

\begin{equation}
\begin{aligned}
	\mathbf{B}(\mathbf{r}) &= \mathbf{B}_\alpha(\mathbf{r}) + \mathbf{B}_\beta(\mathbf{r})
	=\sum_{l=1}^{L_\alpha}\sum_{m=-l}^l \alpha_{lm}\mathbf{B}_{\alpha_{lm}}(\mathbf{r}) + 
	\sum_{l=1}^{L_\beta}\sum_{m=-l}^l \beta_{lm}\mathbf{B}_{\beta_{lm}}(\mathbf{r}),
\end{aligned}
\end{equation}
where $\mathbf{B}_{\alpha_{lm}}(\mathbf{r})$ and $\mathbf{B}_{\beta_{lm}}(\mathbf{r})$
are VSH basis functions
that are perfectly known, representing neuromagnetic and interference field
components, respectively, $\alpha_{lm}$ and $\beta_{lm}$ are the unknown VSH coefficients
that depend on the distribution of the intracranial currents and the
environmental noise sources, respectively. Then, in the notation
of \cite{Taulu2005}, our measurement constitutes a linear operator given by
a VSH matrix $\mathbf{S}$:
\begin{equation}\label{eq:forw}
 \bm{\phi} = \mathbf{Sx},
\end{equation}
where $\bm{\phi}$ is the vector of values measured by the MEG sensor array, $\mathbf{x}$ is the
vector of the VSH coefficients.
\begin{equation*}
	\mathbf{x} = [\alpha_{1,-1},\ldots, \alpha_{L_\alpha, L_\alpha}, \beta_{1,-1},\ldots, \beta_{L_\beta, L_\beta}]^T,
\end{equation*}
and $\mathbf{S}$ is an $m\times n$ matrix determined by the sensor array geometry,
where $m$ is the number of sensors and $n$ is the number
of VSH components. 
%\todo{Is there a special name for the matrix S? \emph{Not really, we
%usually just talk about the SSS matrix, but in that case the matrix typically
%includes both the inside and outside. So, VSH matrix should be fine.}}

Note that the VSH basis allows us to separate the neuronal fields from the
environmental noise. We can write $\mathbf{x}$ as a sum of $\mathbf{x}_\alpha$ 
and $\mathbf{x}_\beta$, containing coefficients for the internal and external 
parts of the VSH expansion:

\begin{equation}
\begin{aligned}
    \mathbf{x} &= \mathbf{x}_\alpha + \mathbf{x}_\beta\\
    \mathbf{x}_\alpha &= \mathbf{I}_\alpha\mathbf{x} \\
    \mathbf{x}_\beta &= \mathbf{I}_\beta\mathbf{x}.
\end{aligned}
\end{equation}
Here $\mathbf{I}_\alpha$ and $\mathbf{I}_\beta$ are diagonal selector matrices that
respectively select only internal or external basis coefficients from $\mathbf{x}$

\begin{equation*}
\mathbf{I}_\alpha = 
\begin{NiceArray}{(cccccc)@{\qquad}l}[nullify-dots]
    1 & & & & & &\\
    & \Ddots & & & & & L_\alpha(L_\alpha +2) \text{ rows}\\
    & & 1 & & & \\
    & & & 0 & & \\
    & & & & \Ddots & & L_\beta(L_\beta +2) \text{ rows}\\
    & & & & &  0  \\
\CodeAfter
    \SubMatrix{.}{4-4}{6-6}{\rbrace}[xshift=3mm]
    \SubMatrix{.}{1-1}{3-6}{\rbrace}[xshift=3mm]
\end{NiceArray}
\end{equation*}

and
\begin{equation*}
\mathbf{I}_\beta = 
\begin{NiceArray}{(cccccc)@{\qquad}l}[nullify-dots]
    0 & & & & & &\\
    & \Ddots & & & & & L_\alpha(L_\alpha +2) \text{ rows}\\
    & & 0 & & & \\
    & & & 1 & & \\
    & & & & \Ddots & & L_\beta(L_\beta +2) \text{ rows}\\
    & & & & &  1  \\
\CodeAfter
    \SubMatrix{.}{4-4}{6-6}{\rbrace}[xshift=3mm]
    \SubMatrix{.}{1-1}{3-6}{\rbrace}[xshift=3mm]
\end{NiceArray}
\end{equation*}

Assuming that $m\geq n$ and the rank of $\mathbf{S}$ is $n$,
we can solve equation \ref{eq:forw} for $\mathbf{x}$:
\begin{equation*}
 \mathbf{x} = \mathbf{S}^\dagger \bm{\phi},
\end{equation*}
where $\mathbf{S}^\dagger$ is the Moore-Penrose pseudoinverse of $\mathbf{S}$.
Now, let us consider some \emph{possible} sensor location (and orientation, as
we assume that our sensor only measures the magnetic field along its
orientation), where we \emph{could have} placed the sensor.
For any possible
location $\mathbf{r} \in V_{\text{samp}}$ and orientation $\mathbf{e}$
($\mathbf{e}$ is a unit vector) the reading of the sensor
$\phi(\mathbf{r}, \mathbf{e})$ would be:

\begin{equation}\label{eq:interp}
 \phi(\mathbf{r}, \mathbf{e}) = \mathbf{s}_{\mathbf{r}, \mathbf{e}}\mathbf{x} = \mathbf{s}_{\mathbf{r}, \mathbf{e}}\mathbf{S}^\dagger \bm{\phi},
\end{equation}
where $\mathbf{s}_{\mathbf{r}, \mathbf{e}}$ is a row vector of length $n$
specifying the values of the VSH components at $(\mathbf{r}, \mathbf{e})$.
Equation \ref{eq:interp} is essentially an interpolation procedure that allows
us to compute the readings of any possible sensor located anywhere in the
sampling volume. 

Now, let us go one step further and say that we want to estimate only the neuronal
component $\phi_\alpha(\mathbf{r}, \mathbf{e})$ of the possible measurement
$\phi(\mathbf{r}, \mathbf{e})$, without the environmental noise. To achieve this
we restrict the interpolation to the inner basis only:
\begin{equation}\label{eq:interp_inner}
 \phi_\alpha(\mathbf{r}, \mathbf{e}) = \mathbf{s}_{\mathbf{r}, \mathbf{e}}\mathbf{x}_\alpha = \mathbf{s}_{\mathbf{r}, \mathbf{e}}\mathbf{I}_\alpha\mathbf{S}^\dagger \bm{\phi}
\end{equation}
Equation \ref{eq:interp_inner} holds exactly if the measurements $\bm{\phi}$
are exact and all the assumptions outlined above (namely, that the first $n$
VSH components capture  all the energy of the magnetic field and $\mathbf{S}$ is
full rank) hold. In this case, it doesn't matter where our sensors are located,
we will always be able to perfectly simulate \emph{any} sensor array
restricted to the sampling volume.\\

However, in reality the sensors are noisy. Thus, instead of reading true values of
magnetic field $\bm{\phi}$, the sensors give us a noisy estimate $\hat{\bm{\phi}}$
\begin{equation}\label{eq:noise}
  \hat{\bm{\phi}} = \bm{\phi} + \bm{\phi}_{\text{noise}}.
\end{equation}
Substituting equation \ref{eq:noise} into \ref{eq:interp_inner} gives us the
noise for the estimation of the $\phi_\alpha$, which we call the interpolation
noise:\footnote{Note that 
computing $\phi_\alpha$ involves not only interpolation, but also external noise
rejection.}
\begin{equation}
	\begin{split}
	\mathbf{s}_{\mathbf{r}, \mathbf{e}}\mathbf{I}_\alpha\mathbf{S}^\dagger \hat{\bm{\phi}} &= \mathbf{s}_{\mathbf{r}, \mathbf{e}}\mathbf{I}_\alpha\mathbf{S}^\dagger (\bm{\phi} + \bm{\phi}_{\text{noise}}) \\
	&= \mathbf{s}_{\mathbf{r}, \mathbf{e}}\mathbf{I}_\alpha\mathbf{S}^\dagger \bm{\phi} + \mathbf{s}_{\mathbf{r}, \mathbf{e}}\mathbf{I}_\alpha\mathbf{S}^\dagger \bm{\phi}_{\text{noise}} \\
	&= \phi_\alpha(\mathbf{r}, \mathbf{e}) + \mathbf{s}_{\mathbf{r}, \mathbf{e}}\mathbf{I}_\alpha\mathbf{S}^\dagger \bm{\phi}_{\text{noise}} \\
	&= \phi_\alpha(\mathbf{r}, \mathbf{e}) + \phi_{\text{noise}}(\mathbf{r}, \mathbf{e}).
	\end{split}
\end{equation}

%\todo{Note that the distribution is a linear function of the sensor noise.
%The mapping from sensor noise to the interpolation noise is a composition of
%two linear mappings, one dependent on the sensor array and the other independent.
%It would be really great if we could remove the independent part and focus on
%the dependent part only. We should talk about this in the discussion.}

If we define our sensor array configuration by a vector $\mathbf{\xi}$ that
contains all sensors' locations and orientations, and observe that
$\mathbf{S^\dagger}$ is a function of $\mathbf{\xi}$, we will see that each
sensor array configuration $\mathbf{\xi}$ yields an interpolation noise
distribution over the sampling volume:
\begin{equation} \label{eq:interp_noise}
	\phi_{\text{noise}}(\mathbf{r}, \mathbf{e}, \mathbf{\xi}) = \mathbf{s}_{\mathbf{r}, \mathbf{e}}\mathbf{I}_\alpha\mathbf{S}^\dagger(\mathbf{\xi}) \bm{\phi}_{\text{noise}}.
\end{equation}
Observe that the term $\mathbf{s}_{\mathbf{r}, \mathbf{e}}\mathbf{I}_\alpha\mathbf{S}^\dagger(\mathbf{\xi})$
in equation \ref{eq:interp_noise} is a row vector of the length $m$ (number of
sensors). We assume that sensor noise $\bm{\phi}_{\text{noise}}$ is a Gaussian,
zero-mean random vector with the elements independent and identically distributed, where each component has variance of $\sigma^2$
\footnote{This is quite a reasonable assumption for real-world MEG devices.
Note, however, that our analysis can be trivially extended to the more general
case where each sensor has a different noise variance.}.
Then $\phi_{\text{noise}}(\mathbf{r}, \mathbf{e}, \mathbf{\xi})$ becomes a
zero-mean Gaussian random variable with variance
$\|\mathbf{s}_{\mathbf{r}, \mathbf{e}}\mathbf{I}_\alpha\mathbf{S}^\dagger(\mathbf{\xi})\|^2\sigma^2$,
where $\|\bullet\|$ denotes the Frobenius norm of a vector. Hence, for each
sensor array configuration $\mathbf{\xi}$ and each location-orientation
pair $(\mathbf{r}, \mathbf{e})$:
\begin{equation}
	\phi_{\text{noise}}(\mathbf{r}, \mathbf{e}, \mathbf{\xi}) \sim \mathcal{N}(0, \sigma_\text{interp}(\mathbf{r}, \mathbf{e}, \mathbf{\xi})^2),
\end{equation}
where
\begin{equation}
	\sigma_\text{interp}(\mathbf{r}, \mathbf{e}, \mathbf{\xi}) = 
	\|\mathbf{s}_{\mathbf{r}, \mathbf{e}}\mathbf{I}_\alpha\mathbf{S}^\dagger(\mathbf{\xi})\|\sigma =
	\|\mathbf{s}_{\mathbf{r}, \mathbf{e}} \mathbf{I}_\alpha \big(\mathbf{S(\mathbf{\xi})}^T\mathbf{S(\mathbf{\xi})}\big)^{-1}\mathbf{S(\mathbf{\xi})}^T\| \sigma.
\end{equation}

%\todo{we should definitely discuss other alternatives, like average}
$\sigma_\text{interp}(\mathbf{r}, \mathbf{e}, \mathbf{\xi})$ describes the
distribution of noise $\phi_{\text{noise}}$ at each location-orientation point
$(\mathbf{r}, \mathbf{e})$. We want to summarize the spatial distribution of
$\sigma_\text{interp}(\mathbf{r}, \mathbf{e}, \mathbf{\xi})$ over the whole
sampling volume with a single value that will serve as a figure-of-merit for
comparing different sensor arrays. There are numerous ways to do this; for
the purpose of this paper we define the figure-of-merit to be the maximum of
$\frac{\sigma_\text{interp}(\mathbf{r}, \mathbf{e}, \mathbf{\xi})}{\sigma}$
over the sampling volume:
\begin{equation}\label{eq:fig-of-merit}
\begin{split}
	q(\mathbf{\xi}) &\triangleq \max_{\substack{\mathbf{r} \in V_{\text{samp}} \\ \|\mathbf{e}\|=1}} \frac{\sigma_\text{interp}(\mathbf{r}, \mathbf{e}, \mathbf{\xi})}{\sigma} \\
	&= \max_{\substack{\mathbf{r} \in V_{\text{samp}} \\ \|\mathbf{e}\|=1}} \|\mathbf{s}_{\mathbf{r}, \mathbf{e}}\mathbf{I}_\alpha\mathbf{S}^\dagger(\mathbf{\xi})\|\\
	&= \max_{\substack{\mathbf{r} \in V_{\text{samp}} \\ \|\mathbf{e}\|=1}} \|\mathbf{s}_{\mathbf{r}, \mathbf{e}} \mathbf{I}_\alpha\big(\mathbf{S(\mathbf{\xi})}^T\mathbf{S(\mathbf{\xi})}\big)^{-1}\mathbf{S(\mathbf{\xi})}^T\|.
\end{split}
\end{equation}

Intuitively, one can think about $q(\mathbf{\xi})$ in the following way:
Assume I have an array $\mathbf{\xi}$ of sensors with additive gaussian noise
of variance $\sigma^2$.
Throughout the array's sampling volume I have a magnetic field that is a
sum of two  components: the brain magnetic field (signal of interest) and the
environmental magnetic field (interference). If my sensor array has the figure-of-merit
value of $q$, it means that I will be able to estimate the signal-of-interest
component of the field anywhere within the sampling volume; my estimate will be noisy 
with additive gaussian noise of standard deviation not worse than $q\sigma$. One can
think of $q$ as the worst-case noise amplification factor; we are going to use the
term \enquote{noise amplification factor} throughout the paper.

Once the figure-of-merit $q(\mathbf{\xi})$ is defined, finding the best sensor
array $\mathbf{\xi}_{opt}$ becomes an optimization problem

\begin{equation}\label{eq:optimize}
	\mathbf{\xi}_{opt} = \argmin_{\mathbf{\xi} \in \mathbf{\Xi} } q(\mathbf{\xi}),
\end{equation}
where $\mathbf{\Xi}$ is the domain of the optimization problem defined by the
sampling volume and the number of sensors $m$ (for the definition of
$\mathbf{\Xi}$ see Equation~\ref{eq:domain_definition}).

In this paper we try to solve equation \ref{eq:optimize} using numerical
nonlinear optimization algorithms. We report the improvement in
$q(\mathbf\xi)$ yielded by the optimization, demonstrate the resulting sensor
geometries, and compare our figure-of-merit to the information-capacity-based
figure-of-merit proposed in the previous works.

\subsection{Channel information capacity of a sensor array}
We wanted to compare the behaviour of our proposed figure-of-merit to some
established metric that has been used by the MEG community. We chose
channel information capacity \parencite{Kemppainen1989} as a reference
metric for such a comparison. Channel information capacity measures the
amount of information (quantified as number of bits per sample) that the
magnetic field as measured by the array conveys about the distribution of
current sources inside the head, under particular assumptions about the source distribution and its statistics.

Under the assumption of spatial white sensor noise with variance $\sigma^2$ and a Gaussian source distribution with a covariance matrix $\mathbf{\Sigma}$, the information capacity can be calculated as \parencite{Kemppainen1989}
\begin{equation}
    I = \frac{1}{2} \sum_i \log_2 (P_i + 1) =  \frac{1}{2} \sum_i \log_2 \left(\frac{\lambda_i^2}{\sigma^2} + 1\right),
\end{equation}
where $P_i$ are the SNRs of the orthogonal channels defined by the eigencomponents and the eigenvalues $\lambda_i$ of the covariance matrix $\mathbf{L} \mathbf{\Sigma} \mathbf{L}^T$, where $\mathbf{L}$ is the lead-field matrix representing the measured magnetic fields produced by the sources.
%(e.g.
%that the sources are point-like \todo{is this a good word?} current
%dipoles uniformely distributed across a spherical volume, etc.).

\subsection{Implementation details}
We did all the computations in Python 3 programming language using popular
libraries for scientific computing and visualization such as SciPy \parencite{virtanen2020scipy}, NumPy \parencite{harris2020array},
and mayavi \parencite{ramachandran2011mayavi}. All the source code used for the simulations described in this
paper is available from GitHub\footnote{\url{https://github.com/andreyzhd/MEGSim}}
under the GNU General Public License \parencite{gplv3}.

\subsubsection*{VSH computation}
We computed $\mathbf{S}(\mathbf{\xi})$ and $\mathbf{s}_{\mathbf{r}, \mathbf{e}}$ using the implementation of VSHs in
 MNE-Python \parencite{Gramfort2013}. We used $L_\alpha=10$ and
$L_\beta=3$ for the VSH expansion, which resulted in 135 components in the expansion.

\subsubsection*{Approximating spatial distribution of interpolation noise}
Theoretically, $q(\mathbf\xi)$ is defined as a maximum of a continuous
function $\|\mathbf{s}_{\mathbf{r}, \mathbf{e}}\mathbf{I}_\alpha\mathbf{S}^\dagger(\mathbf{\xi})\|$
over a bounded domain
$\{(\mathbf{r}, \mathbf{e}) | \mathbf{r} \in V_\text{samp}, \|\mathbf{e}\|=1\} = \{\mathbf{r} | \mathbf{r} \in V_\text{samp}\} \times \{\mathbf{e} | \|\mathbf{e}\|=1\}$
(see equation \ref{eq:fig-of-merit}). In practice, we approximated the
continuous domain $\{\mathbf{r} | \mathbf{r} \in V_\text{samp}\}$ by a
dense discrete grid of 2500 points for the 3D and 1000 points for the 2D
sampling volumes.

For the 2D sampling volume, the 1000 points are approximately uniformly spread
across the helmet surface \footnote{We distribute the points on the helmet
surface are using a variation of the \enquote{golden ratio} algorithm
for approximately evenly distributing points on a spherical surface.}
(see figure~\ref{fig:sampl_vol} left). Helmet surface being approximately 0.25 m$^2$,
the resulting density of the sampling locations is about 1 location per 0.00025 m$^2$.

For the 3D sampling volume, the sampling grid comprises 5 concentric shells, each
shell similar to the 2D volume described above. The shells radii are uniformly
distributed on the interval 0.15 -- 0.25 m, making the radial spacing between two
neighboring shells 0.02 m. Each shell has 500 sampling locations uniformly
distributed across it, for the outermost shell this leads to the density of about
1 sampling location per 0.001 m$^2$.

\subsubsection*{Initialization of the optimization procedure}
The optimization procedure is initialized with a uniformly-spaced radial
sensor configuration (see section \nameref{sec:uniformely_spaced_arrays} for more details).
For the 3D array optimization procedure, we try three different initial conditions
corresponding the radii $R=0.15$ m, $R=0.2$ m, and $R=0.25$ m for
the initial sensor array configuration.

\subsubsection*{Optimization procedure}
We evaluated several general-purpose nonlinear optimization algorithms: Basin-Hopping,
Differential Evolution, and Dual Annealing. Of these, the Dual Annealing  
demonstrated the best performance, so we used it for all the work
described in the paper.

The Dual Annealing, as implemented by the scipy.optimize.dual\_annealing function of the
scipy toolbox, is a stochastic optimization algorithm derived from the Generalized
Simulated Annealing \parencite{Xiang1997}. This method combines the Classical Simulated
Annealing (CSA) with the Fast Simulated Annealing (FSA) algorithms augmented by a local
search on accepted locations.

We used the default values for the maximum number of global iterations (1000) and the
limit for the number of objective function calls $(10^7)$. These parameters resulted in a
optimization run lasting 3--5 days on a typical desktop computer.

It is important to note, that Dual Annealing is an optimization procedure over a continuous
parameter space: the parameter variables are not restricted to a set of possible discrete values.
The only constraint that we used during the optimization process was the requirement that all
the sensors should be inside the sampling volume.

%\todo{add citations here}

\subsubsection*{Channel information capacity computation}
We used 1,000 random current dipoles to compute the channel information capacity. Each
dipole's location was randomly chosen from a uniform distribution from a spherical
volume of radius 0.07 m centered at the origin. Each dipole's orientation was
randomly chosen from a uniform distribution on a sphere. The total dipole moment
(root-sum-squared across all the dipoles) was $2\cdot 10^{-8}$ A$\cdot$m and the standard deviation
of the sensor noise was $10^{-14}$ T.

\subsection{Computational Experiments}
In this paper we report the results of three computational experiments.

\subsubsection*{Investigation of Uniformly-Spaced Radial Arrays}
As a uniformly-shaped radial sensor array is a function of its radius $R$
and the number of sensors $N$, in our first computational experiment we study the
behavior of the array's
noise amplification factor as a function of these two parameters.

\subsubsection*{Array Optimization based on a 3D Sampling Volume}
In the second experiment we try to find an optimal (w.r.t. the noise amplification
factor) design for a sensor array of 240 sensors within a 3D sampling volume.
We investigate the stabilty of the optimization procedure w.r.t. the starting condition
by running multiple experiments with different initial conditions.

Additionally, we investigate the behavior of the optimization procedure for
different orders of the expansion of the VSH basis.

\subsubsection*{Array Optimization based on an anatomically-constrained 3D Sampling Volume}
In the third experiment we perform a single optimization run using an
anatomically-constrained 3D Sampling Volume.

\subsubsection*{Array Optimization based on a 2D Sampling Volume}
In this experiment we repeat the optimization experiment we performed on a 3D
sampling volume, but this time on a 2D sampling volume of radius 15 cm. Note
that 15 cm is the inner radius of the 3D sampling volume; however 2D volume-based
optimization is not the same as the 3D optimization with sensor
locations restricted to a 2D surface. The two procedures use different fitness
functions, since they have different sampling volumes.

%% file: results.tex
\section{Results}
\label{results}
\subsection{Investigation of Uniformly-Spaced Radial Arrays}
\begin{figure}[ht]
    \centering
    \includegraphics[width=\textwidth]{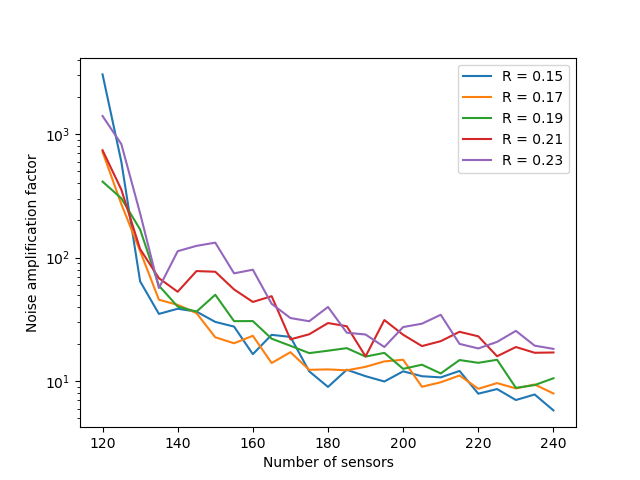}
    \caption{Behaviour of the noise amplification factor for uniformly-spaced
    radial arrays as a function of the number of sensors and array radius $R$.}
    \label{fig:reg_nsens}
\end{figure}

\begin{figure}[ht]
    \centering
    \includegraphics[width=\textwidth]{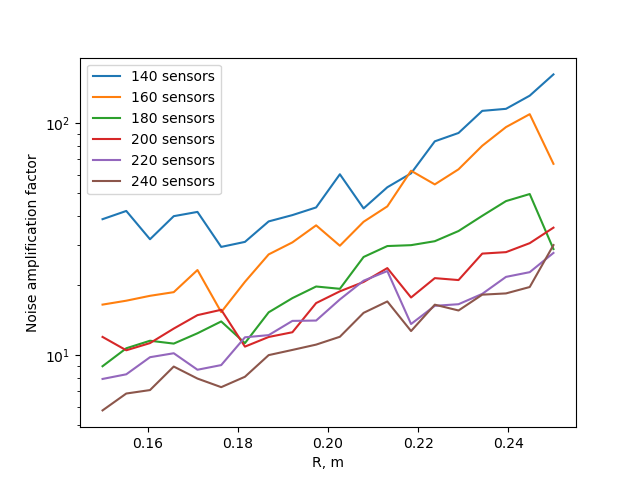}
    \caption{Behaviour of the noise amplification factor for uniformly-spaced
    radial arrays as a function of the number of sensors and array radius $R$.}
    \label{fig:reg_R}
\end{figure}

Figures \ref{fig:reg_nsens} and \ref{fig:reg_R} show the behaviour of the noise amplification factor for uniformly-spaced
    radial arrays as a function of array radius $R$ and the number of sensors. The noise amplification factor was calculated based on equation~\ref{eq:fig-of-merit}. From figure \ref{fig:reg_nsens}, we see that when the number of sensors is doubled from 120 to 240, the noise amplification factor shows a reduction of roughly two orders of magnitude. Figure \ref{fig:reg_R} shows that the noise amplification factor improves when the array radius $R$ decreases.
    
\subsection{Array Optimization based on a 3D Sampling Volume}

\begin{figure}[ht]
    \centering
    \includegraphics[width=\textwidth]{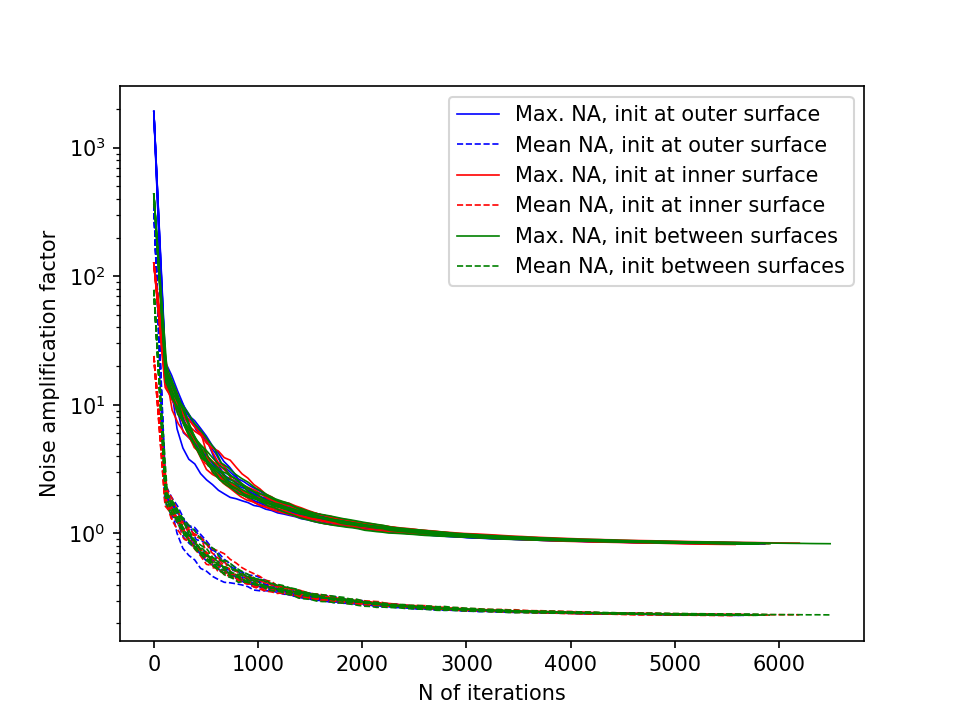}
    \caption{Noise amplification factor (NA) as a function of iteration during a 3D sampling volume-based optimization procedure. The optimization was repeated for different initial sensor locations: sensors on outer surface, inner surface, and halfway between the surfaces. N=8 runs are plotted for each of these conditions. The solid and dashed lines indicate the maximum and mean noise amplification, respectively.}
    \label{fig:noise_opt}
\end{figure}

\begin{figure}[ht]
    \centering
    \includegraphics[width=\textwidth]{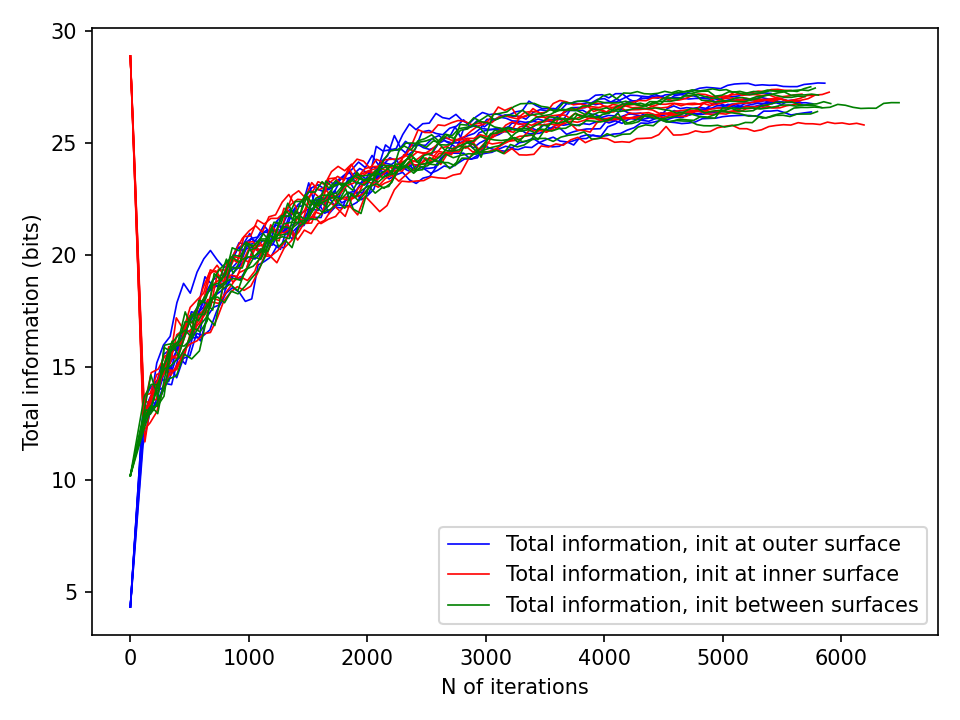}
    \caption{Total information as a function of iteration during a 3D sampling volume-based optimization procedure. The optimization was repeated for different initial sensor locations: sensor on outer surface, inner surface, and halfway between the surfaces. N=8 runs are plotted for each of these conditions.}
    \label{fig:inf_opt}
\end{figure}

Figures \ref{fig:noise_opt} and \ref{fig:inf_opt} depict the behavior of the sensor array's noise amplification factor and channel information capacity during a 3D sampling-volume-based optimization procedure. Both the noise amplification factor and the channel information capacity improve as more iterations are performed. The maximum and average noise amplification factors saturate approximately at values 1.0 and 0.2, respectively, while the channel information capacity reaches a value of approximately 30 bits per sample. The optimization was repeated with different initial sensor locations. In general, the results are quite consistent between runs, and the initial locations do not have a significant effect on the final optimization result.

\begin{figure}[ht]
    \centering
    \begin{tabular}{cccc}
    \small{iteration 0} &
    \small{iteration 2967} &
    \small{iteration 4450} &
    \small{iteration 5934} \\
    \includegraphics[width=0.25\textwidth]{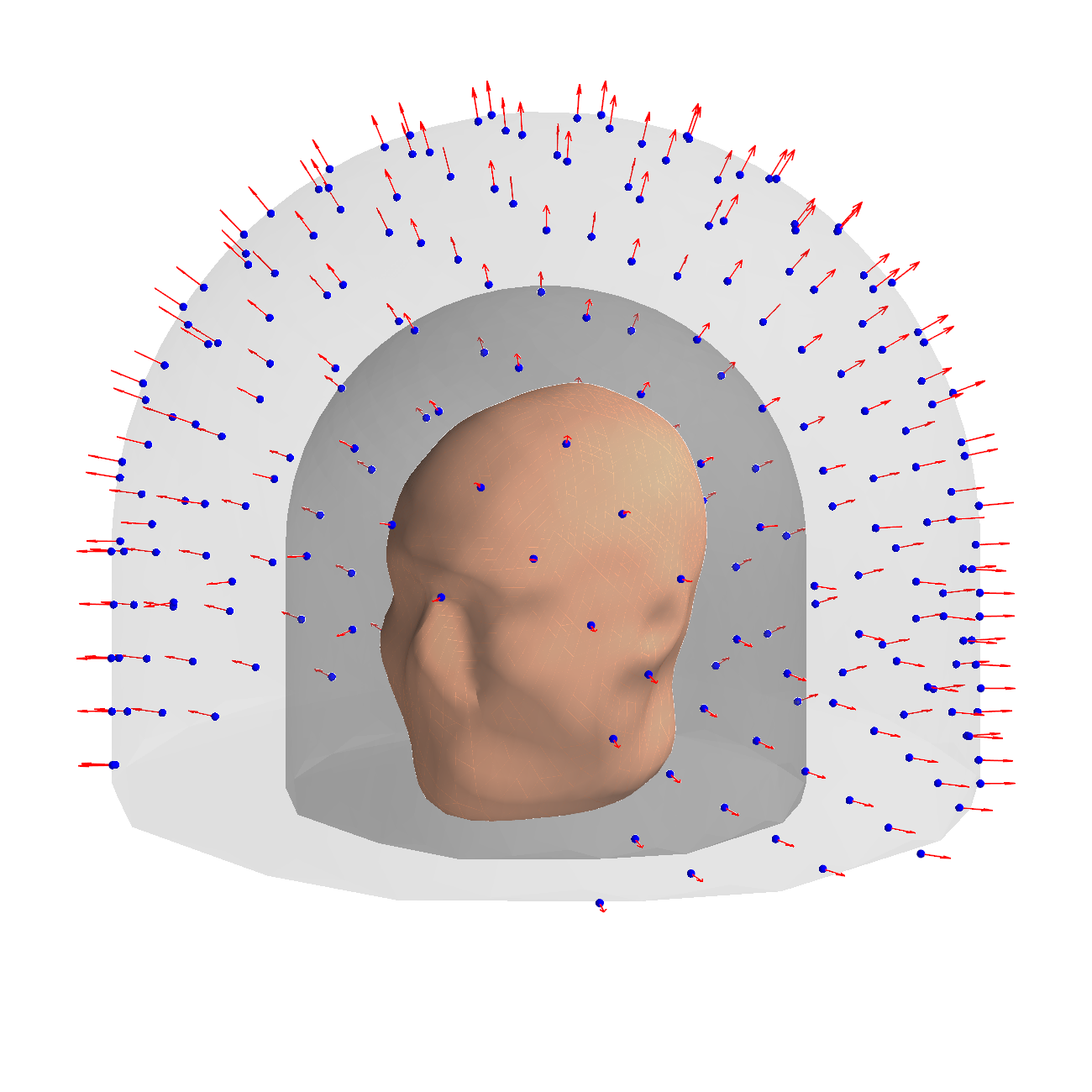} &
    \includegraphics[width=0.25\textwidth]{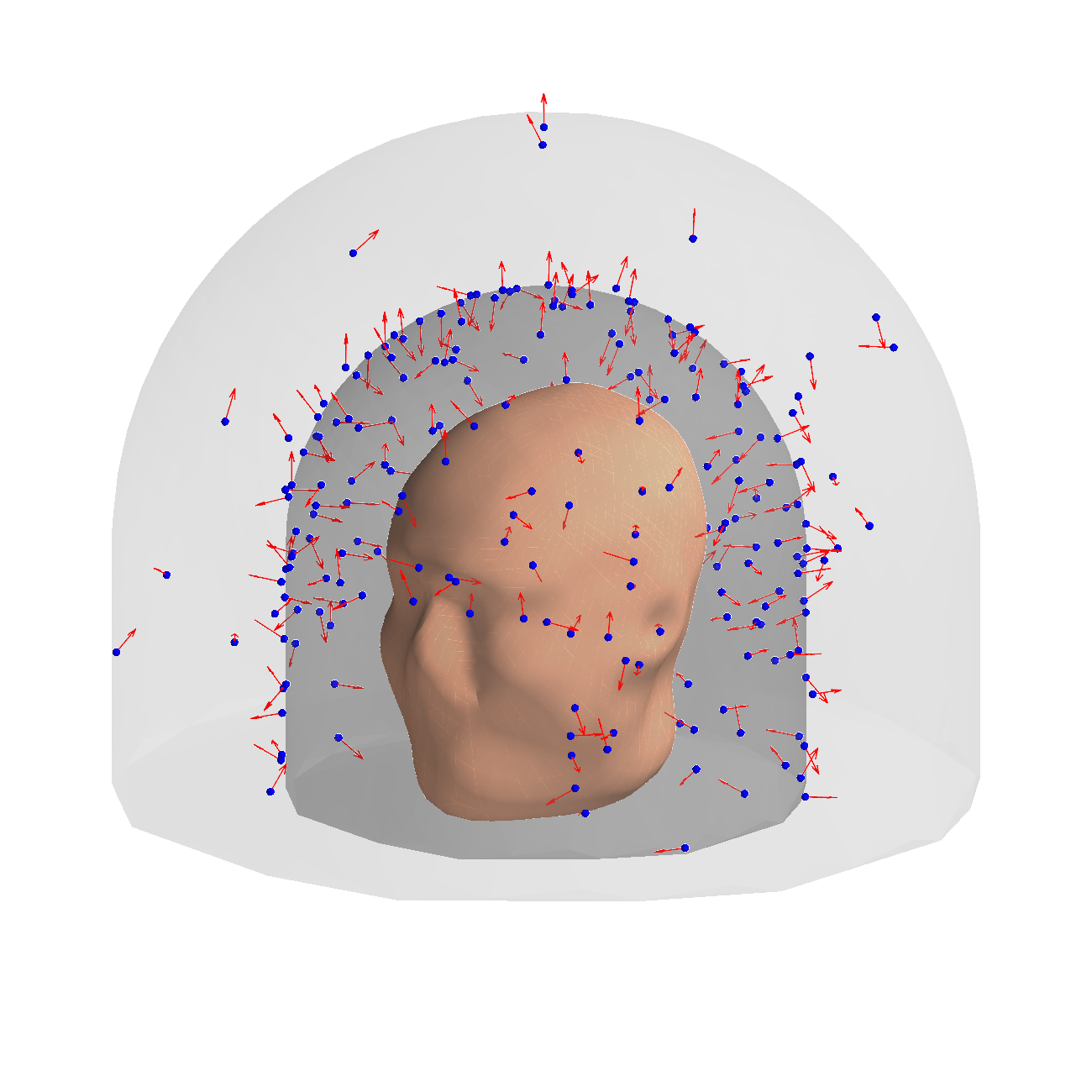} &
    \includegraphics[width=0.25\textwidth]{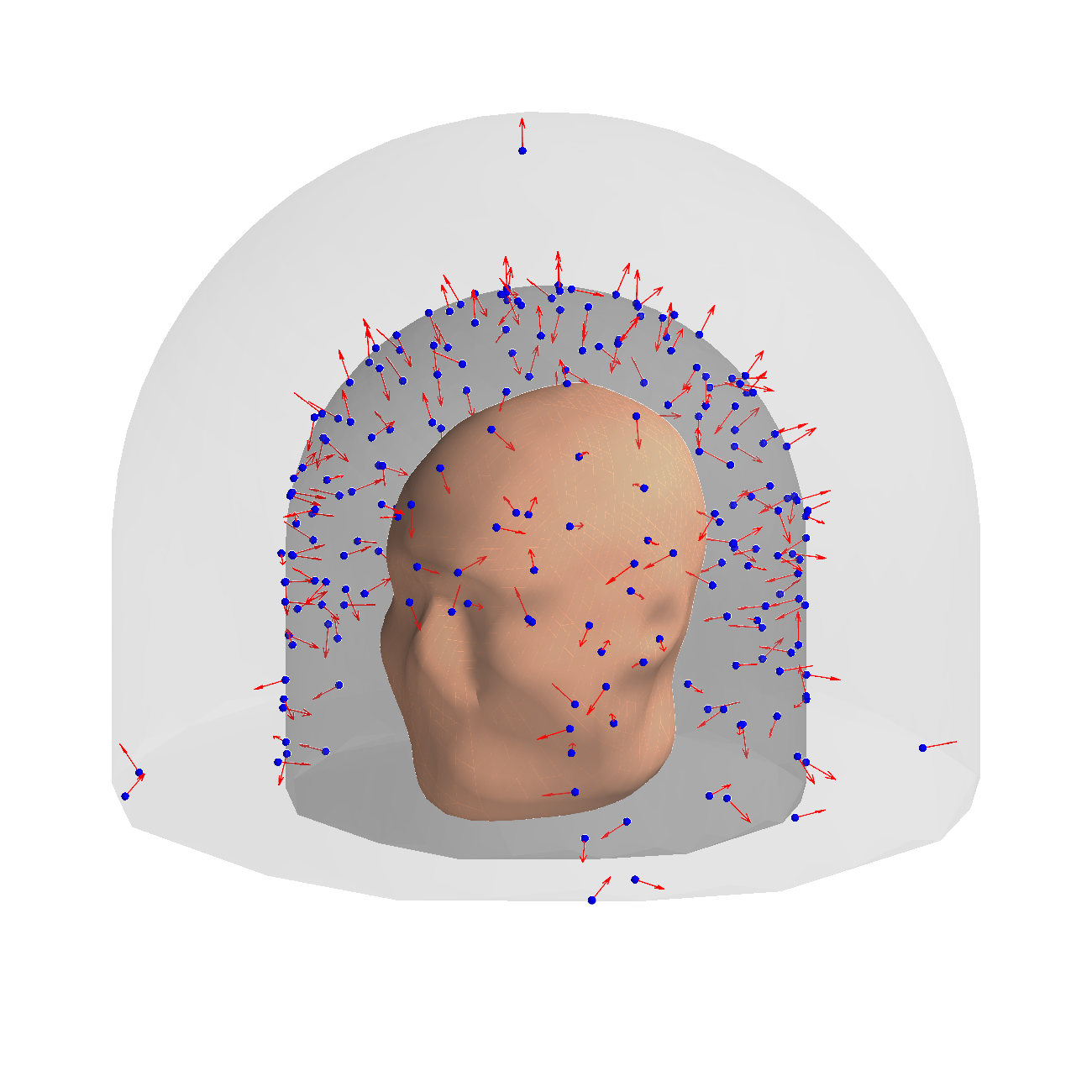} &
    \includegraphics[width=0.25\textwidth]{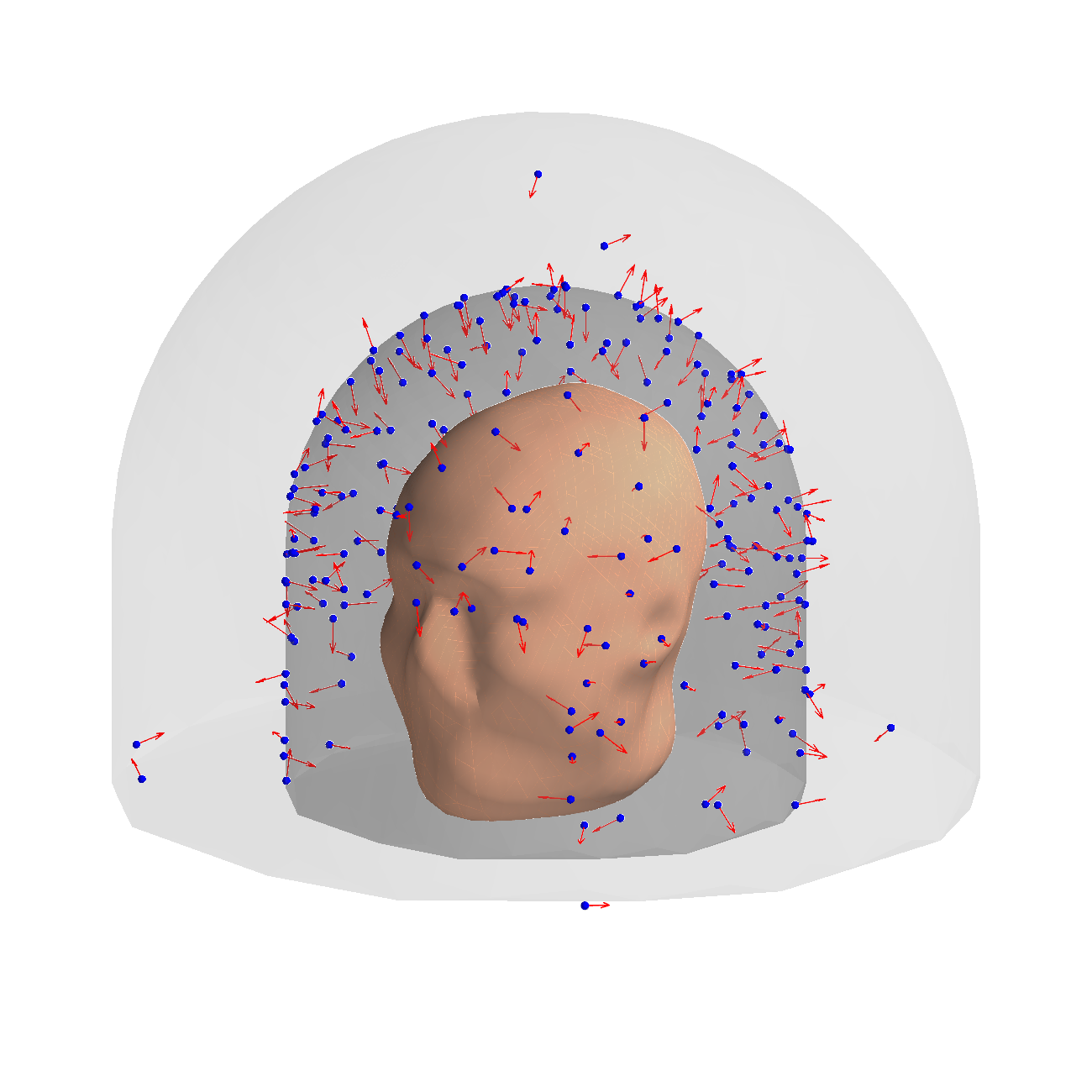} 
    \end{tabular}
    \caption{Progression of the sensor arrangement during the optimization for a
    3D sampling volume.}\label{fig:geom_opt_thick}
\end{figure}

Figure~\ref{fig:geom_opt_thick} depicts the evolution of the sensor array geometry during one optimization run, where the initial location of the sensors is on the outer surface. As the algorithm progresses, the sensors mostly migrate to the inner surface. On the average, about 6 sensors out of 240 remained close to the outer surface.

\begin{figure}[ht]
    \centering
    \includegraphics[width=\textwidth]{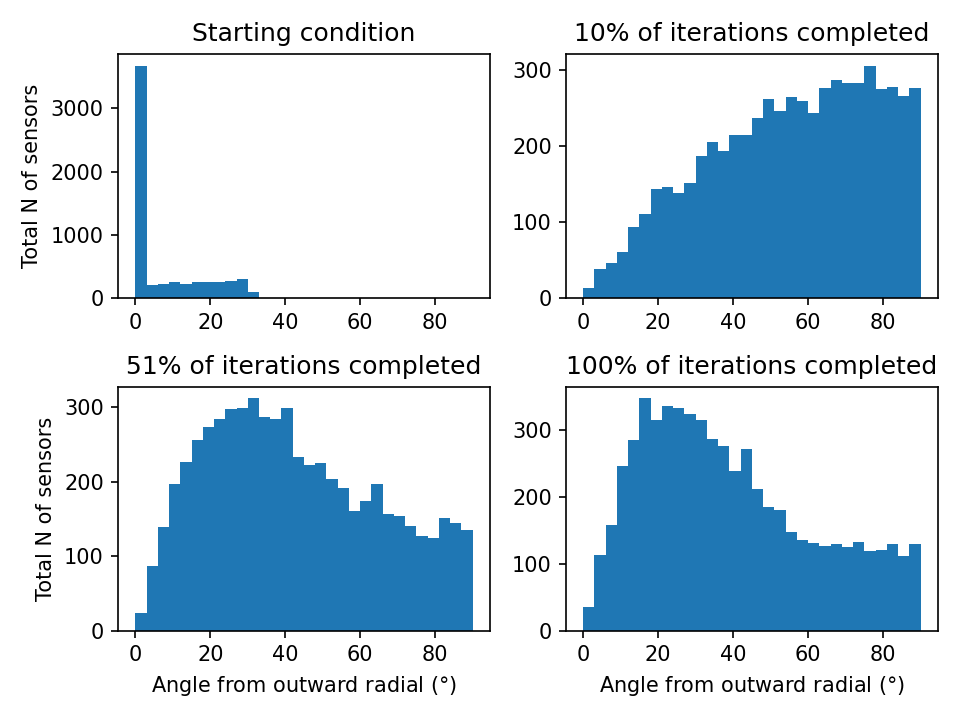} 
    \caption{Distribution of sensor orientations during optimization. Data from
    N=25 runs with the sensors starting at the outer surface are combined in the plot,
    for a total of 6000 sensors. "Outward radial" refers to the radial normal of
    the spherical coordinate system, with the origin at the center of the sensor
    helmet.}\label{fig:angle_distribution}
\end{figure}

Figure~\ref{fig:angle_distribution} illustrates the distribution of sensor orientations during optimization. In the initial condition, the orientations are mostly aligned with the radial normal of the spherical coordinate system; the alignment is not exact due to the helmet-like shape of the sensor array. At early stages of optimization, the tangential directions start to dominate. At convergence, the sensors have mixed orientations, with a majority of them being oriented more towards the radial direction.

\begin{figure}[ht]
    \centering
    \includegraphics[width=\textwidth]{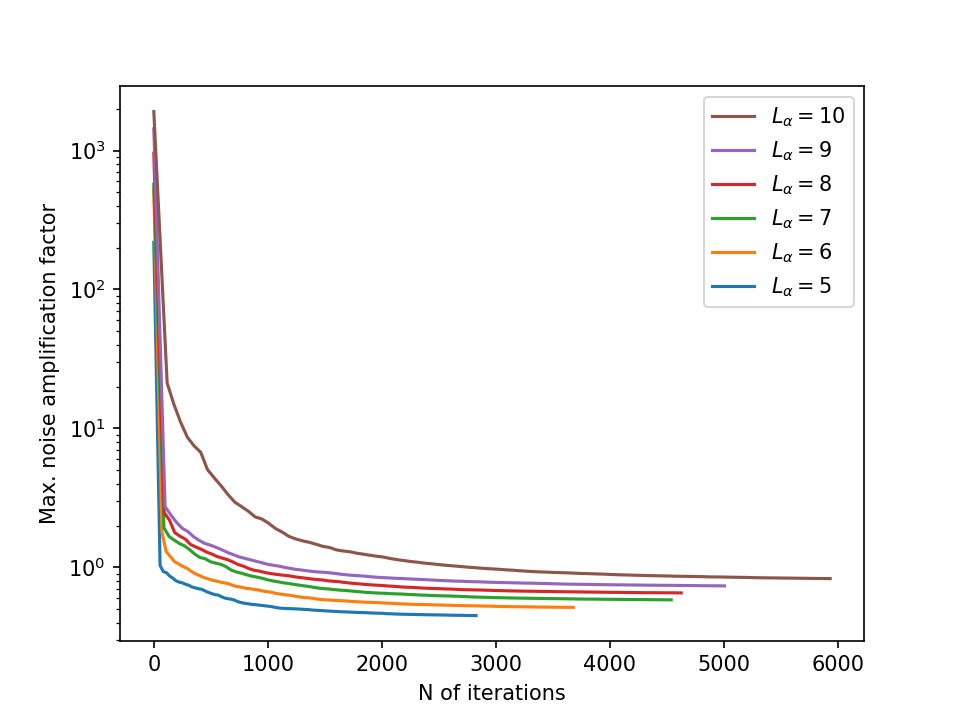}
    \caption{Maximum noise amplification factor (NA) as a function of iteration during a 3D sampling volume-based optimization procedure for different values of $L_\alpha$, with $L_\beta$ = 3.}
    \label{fig:lin_opt}
\end{figure}

Finally, figure \ref{fig:lin_opt} illustrates the dependence on the method on the selected VSH degree cutoff. The optimization was repeated for different values of $L_\alpha$. Using a lower VSH degree cutoff results in faster convergence and lower overall noise amplification for the sensor array.

% Important! The command below reduces column margins from here onwards!
\setlength\tabcolsep{0pt} 

\subsection{Array Optimization based on an anatomically-constrained 3D Sampling Volume}
\begin{figure}[ht]
    \centering
    \includegraphics[width=\textwidth]{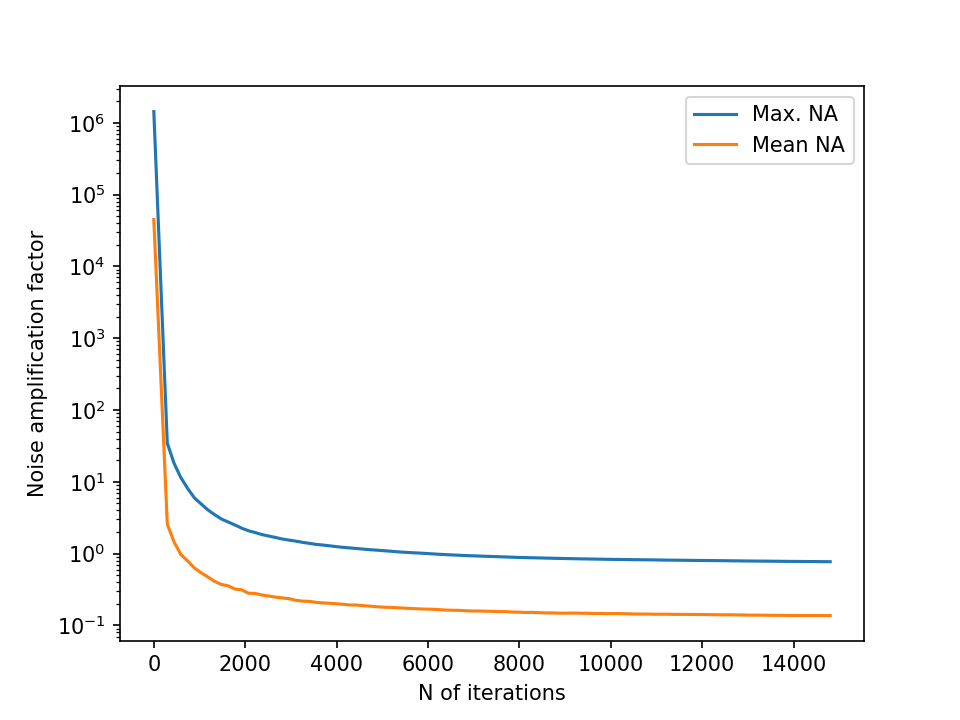}
    \caption{Noise amplification factor (NA) as a function of iteration for optimization on anatomically-constrained 3D Sampling Volume.}
    \label{fig:noise_opt_scalp}
\end{figure}

\begin{figure}[ht]
    \centering
    \includegraphics[width=\textwidth]{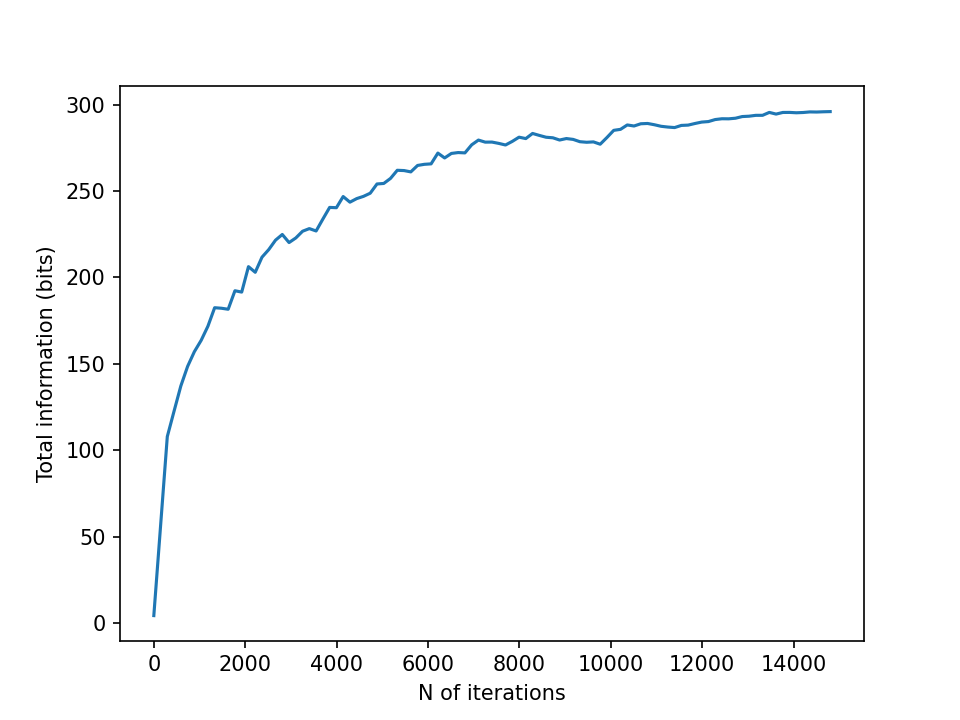}
    \caption{Channel information capacity as a function of iteration for optimization on anatomically-constrained 3D Sampling Volume.}
    \label{fig:inf_opt_scalp}
\end{figure}

Figures \ref{fig:noise_opt_scalp} and \ref{fig:inf_opt_scalp} depict the behavior of the
sensor array's noise amplification factor and channel information capacity during an
optimization procedure for the anatomically-constrained 3D sampling volume. The optimization
procedure generally behaves very similarly to that of the regular 3D sampling volume. One
major difference is that the anatomically-constrained version attains much higher channel
information capacity, which is to be expected, considering the fact that it can position
sensors much closer to the sources of the signal inside the head.

\begin{figure}[ht]
    \centering
    \begin{tabular}{cccc}
    \small{iteration 0} &
    \small{iteration 7399} &
    \small{iteration 11099} &
    \small{iteration 14799} \\
    \includegraphics[width=0.25\textwidth]{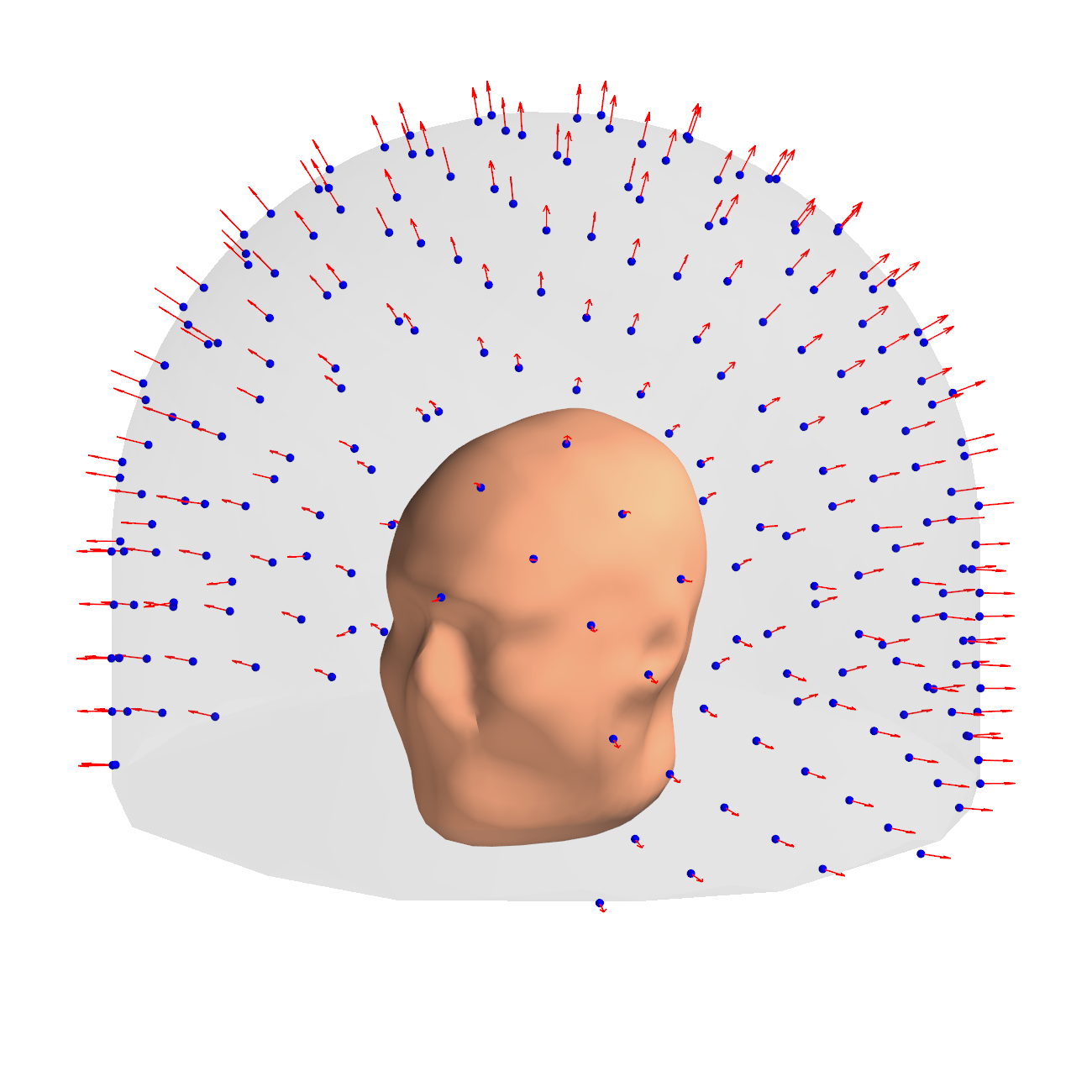} &
    \includegraphics[width=0.25\textwidth]{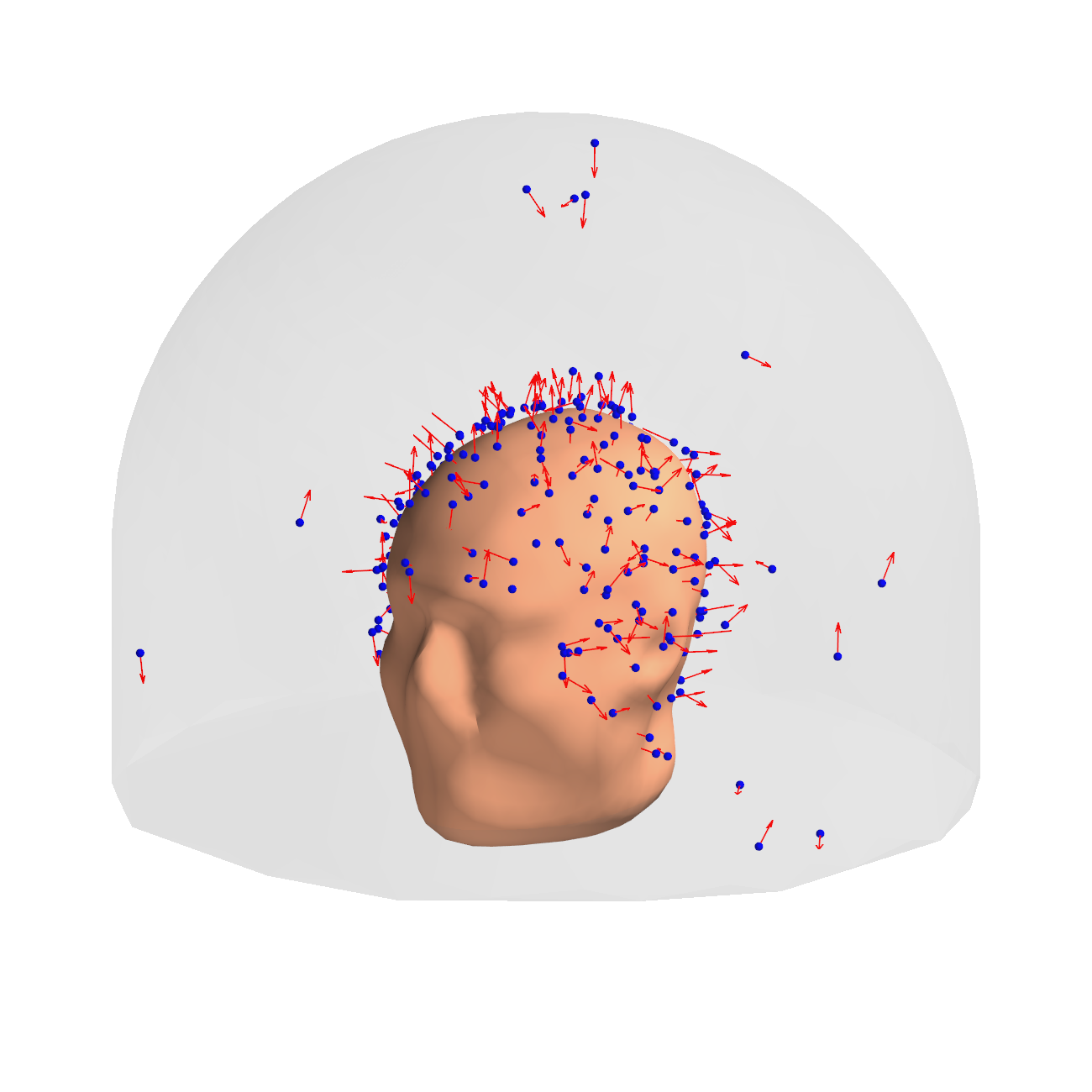} &
    \includegraphics[width=0.25\textwidth]{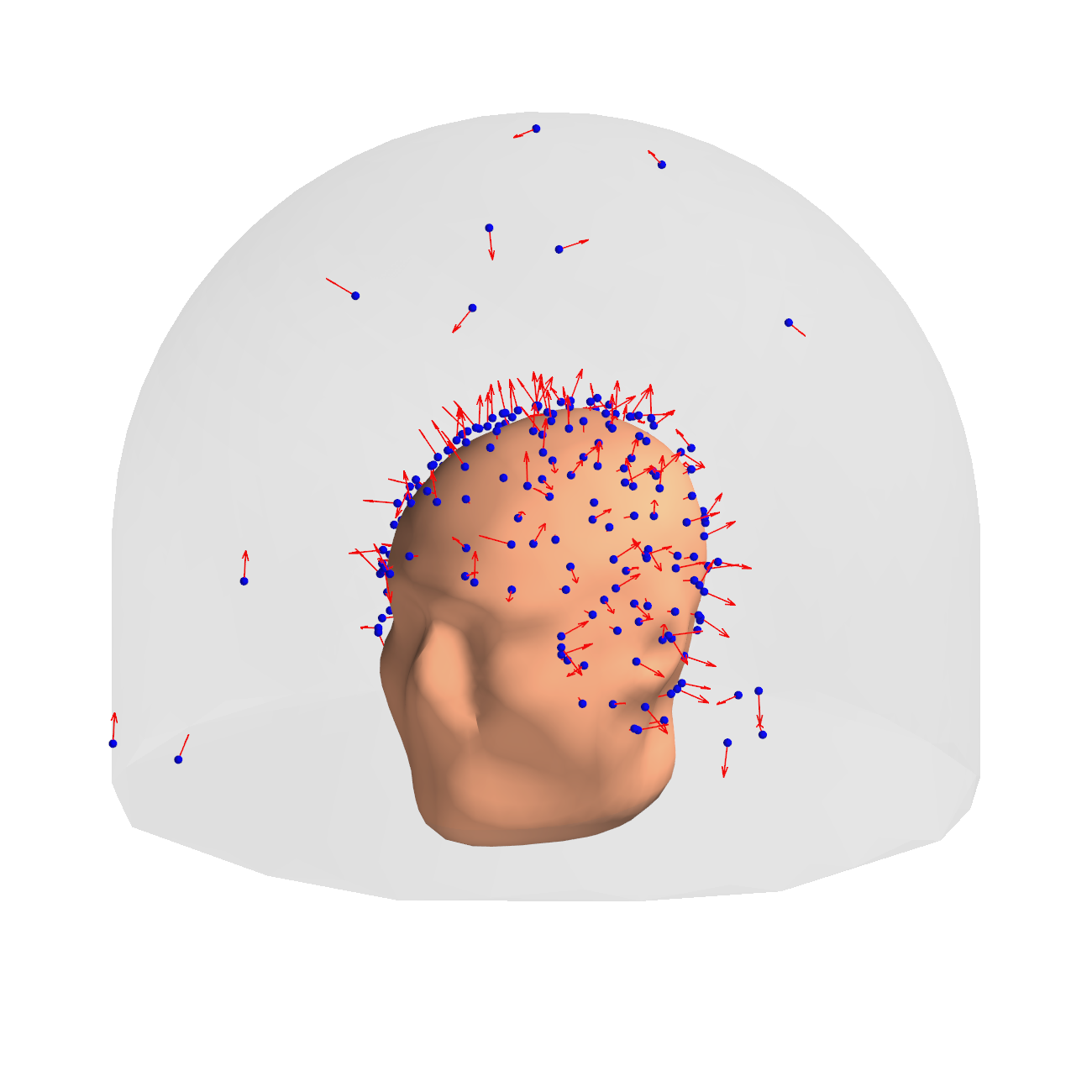} &
    \includegraphics[width=0.25\textwidth]{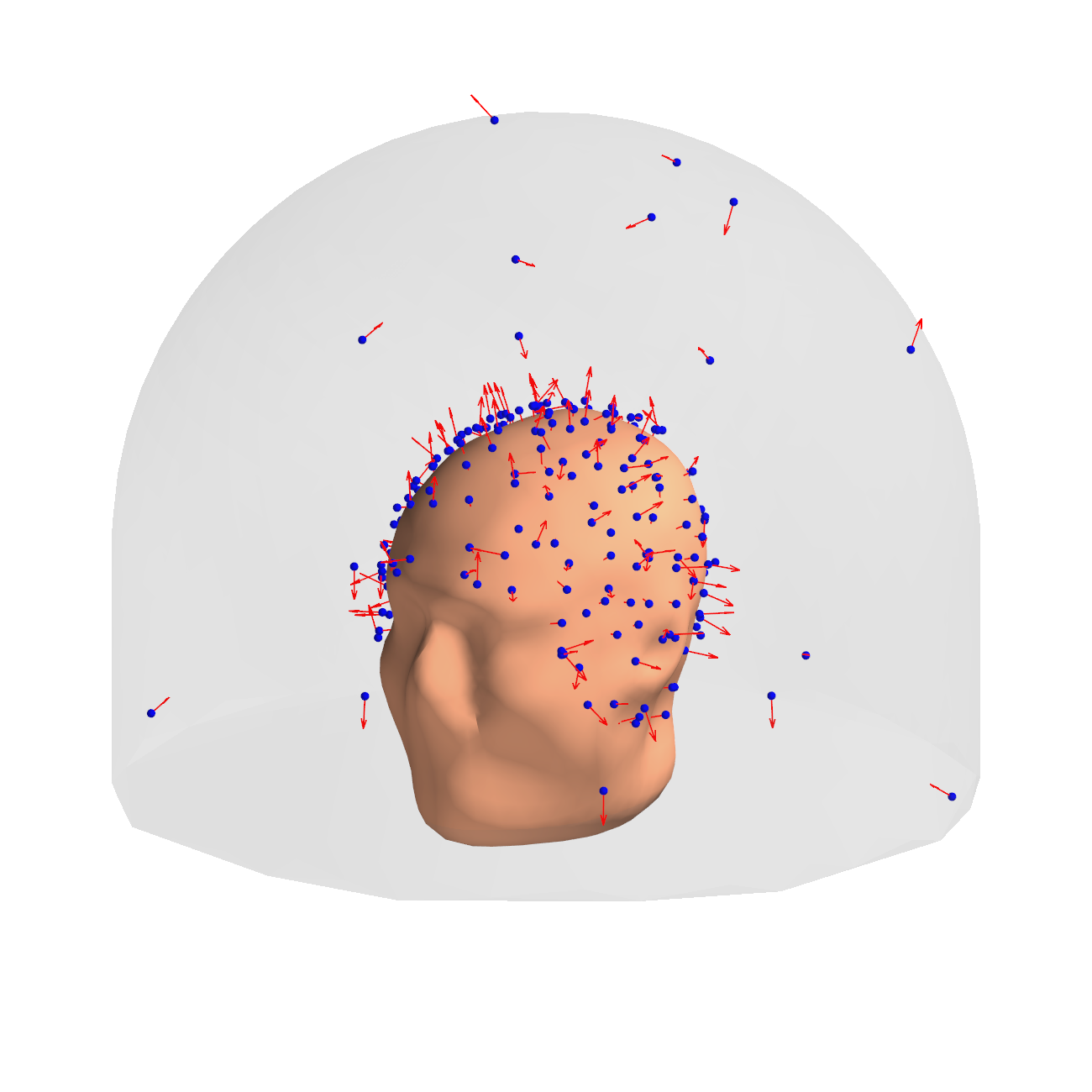} 
    \end{tabular}
    \caption{Progression of the sensor arrangement during the optimization for an
    anatomically-constrained 3D sampling volume.}\label{fig:geom_opt_scalp}
\end{figure}

Figure~\ref{fig:geom_opt_scalp} depicts the evolution of the sensor array
geometry during an optimization run. This too is qualitatively similar
to the results for the regular 3D array.

\subsection{Array Optimization based on a 2D Sampling Volume}
\begin{figure}[ht]
    \centering
    \includegraphics[width=\textwidth]{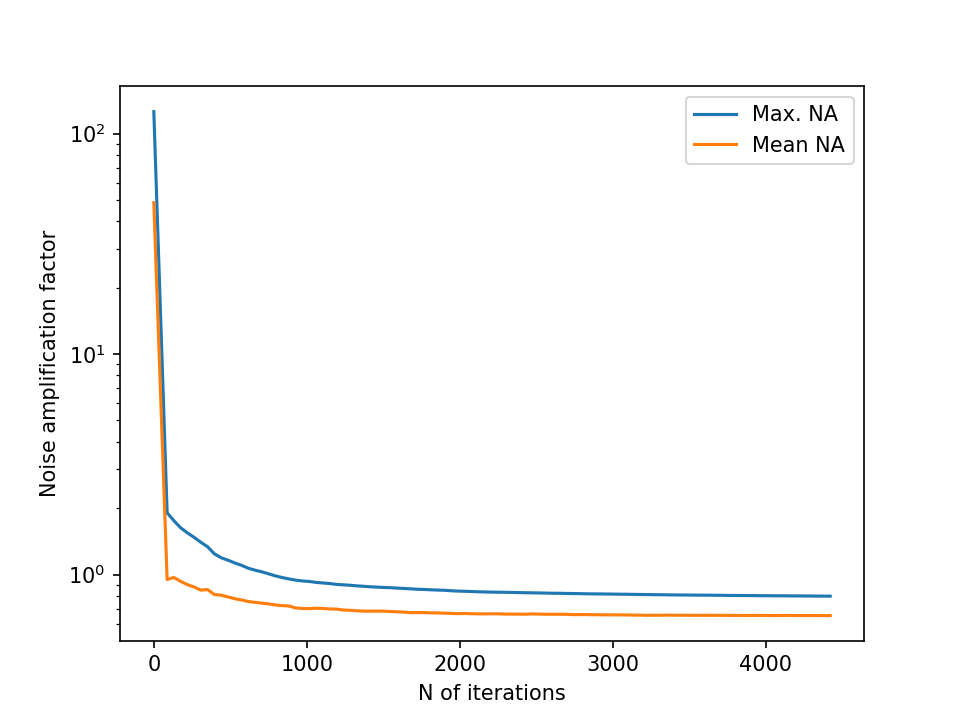}
    \caption{Noise amplification factor (NA) as a function of iteration for a 2D sampling volume-based optimization procedure.}
    \label{fig:noise_opt_thin}
\end{figure}

\begin{figure}[ht]
    \centering
    \includegraphics[width=\textwidth]{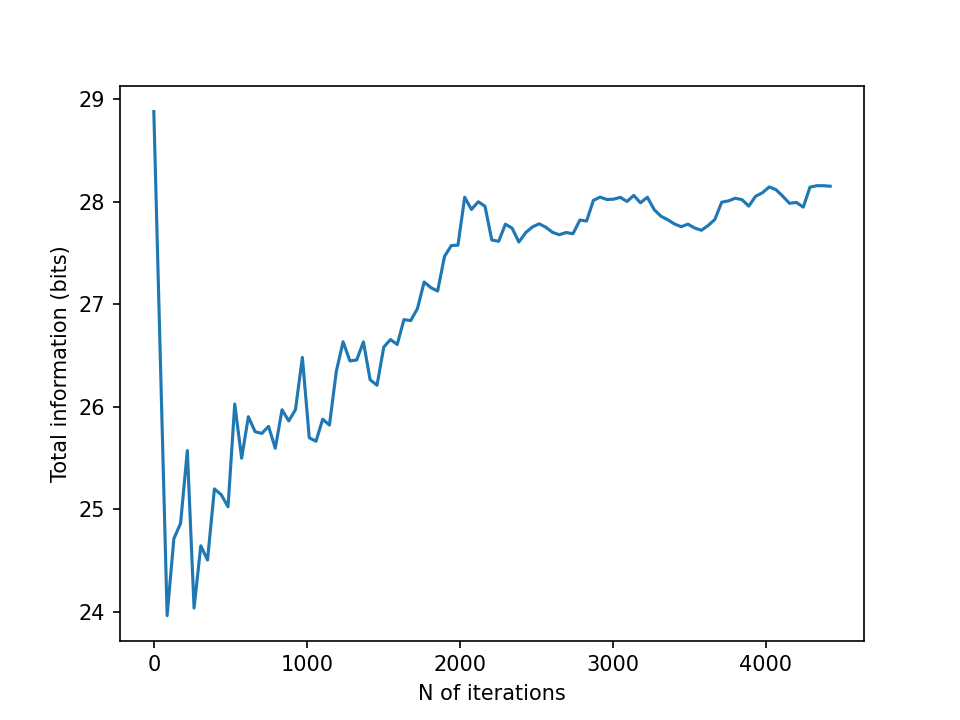}
    \caption{Channel information capacity as a function of iteration for a 2D sampling volume-based optimization procedure.}
    \label{fig:inf_opt_thin}
\end{figure}

\begin{figure}[ht]
    \centering
    \begin{tabular}{cccc}
    \small{iteration 0} &
    \small{iteration 2210} &
    \small{iteration 3315} &
    \small{iteration 4421} \\
    \includegraphics[width=0.25\textwidth]{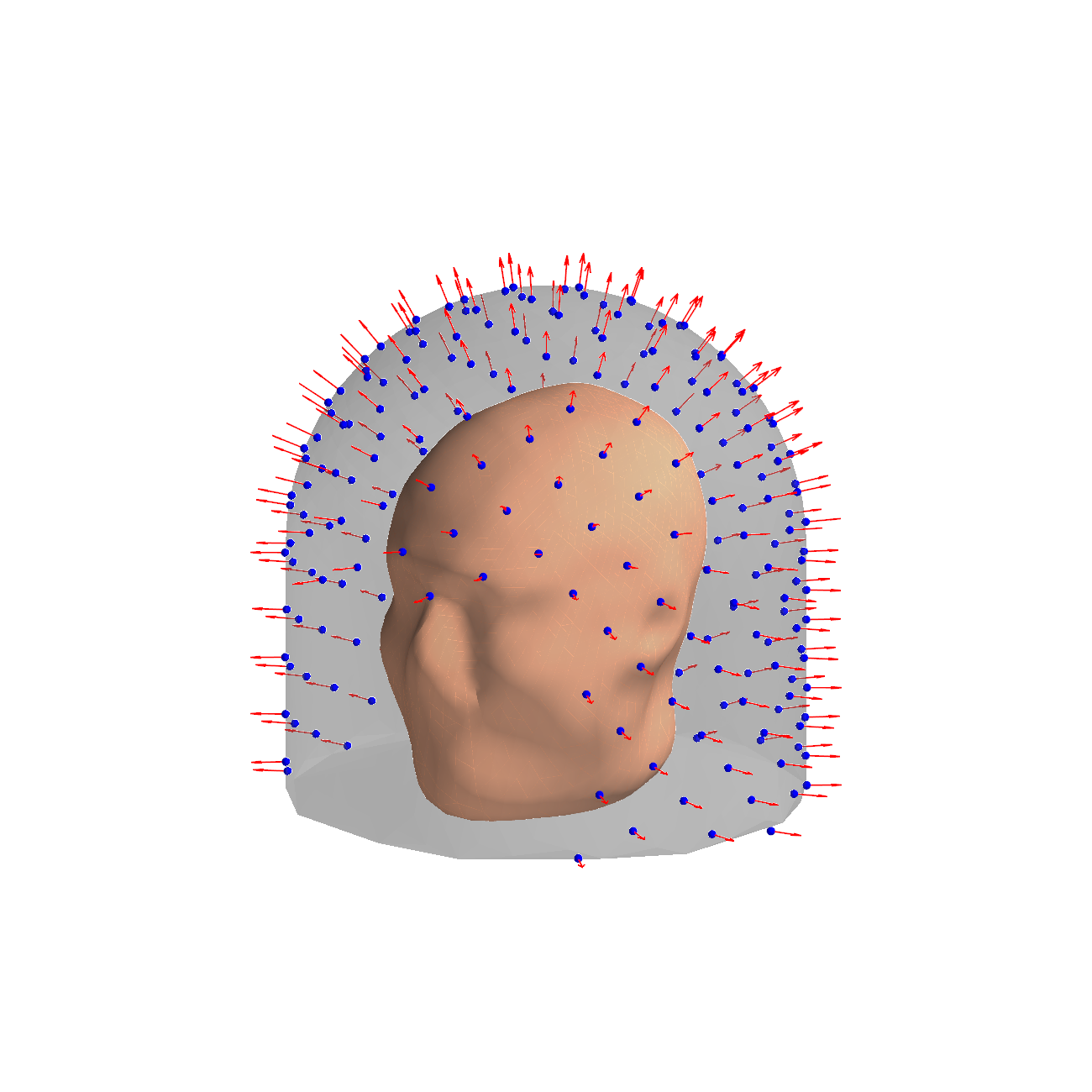} &
    \includegraphics[width=0.25\textwidth]{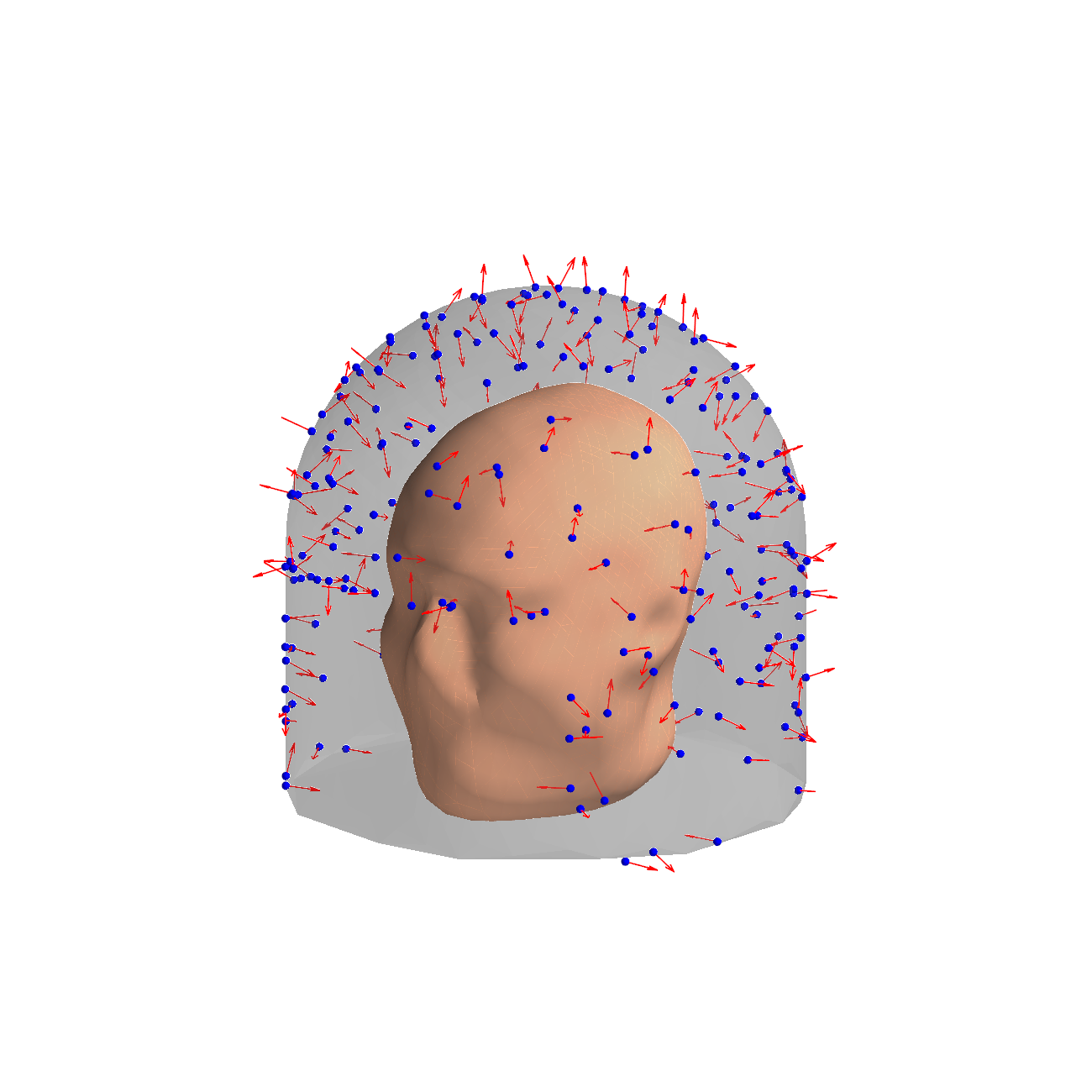} &
    \includegraphics[width=0.25\textwidth]{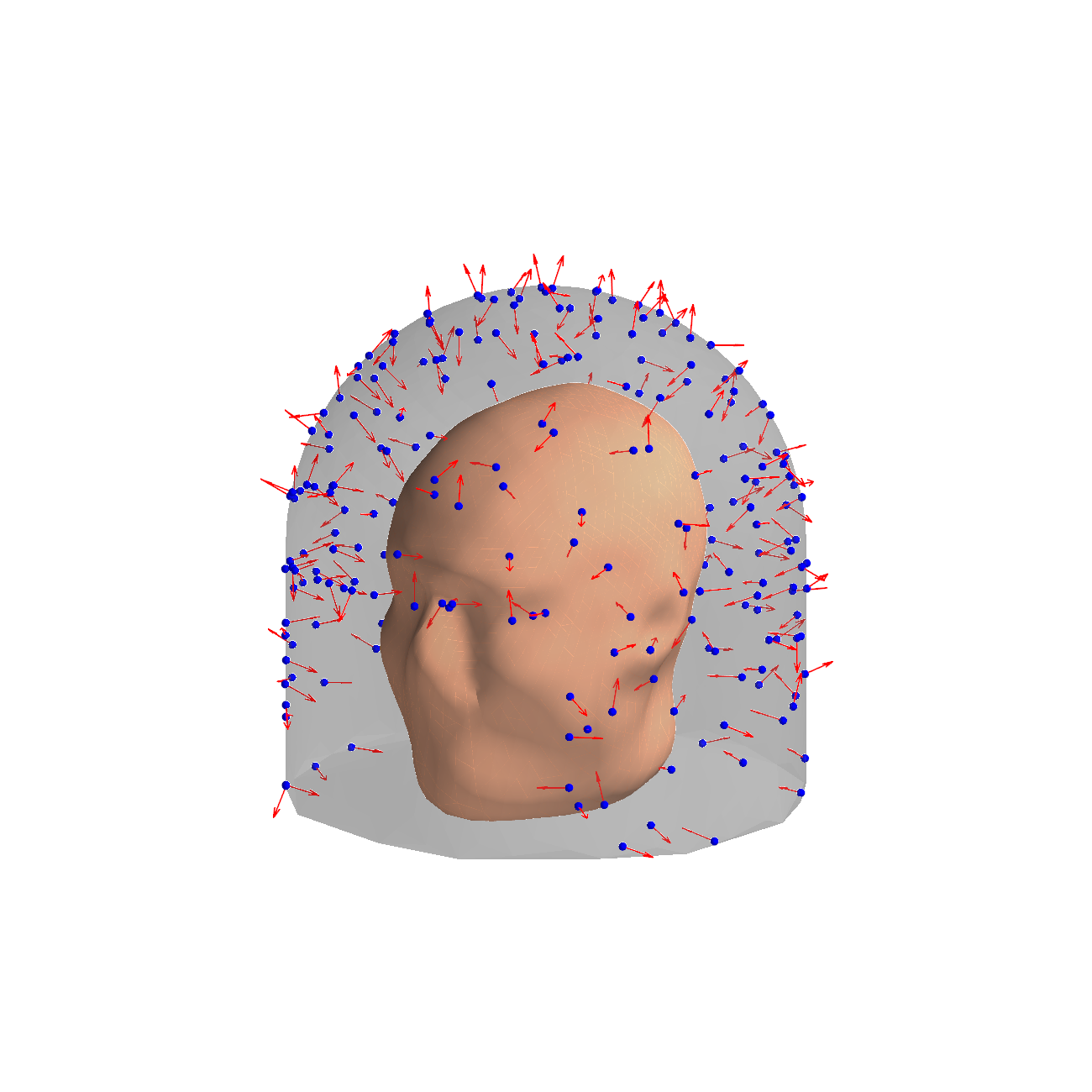} &
    \includegraphics[width=0.25\textwidth]{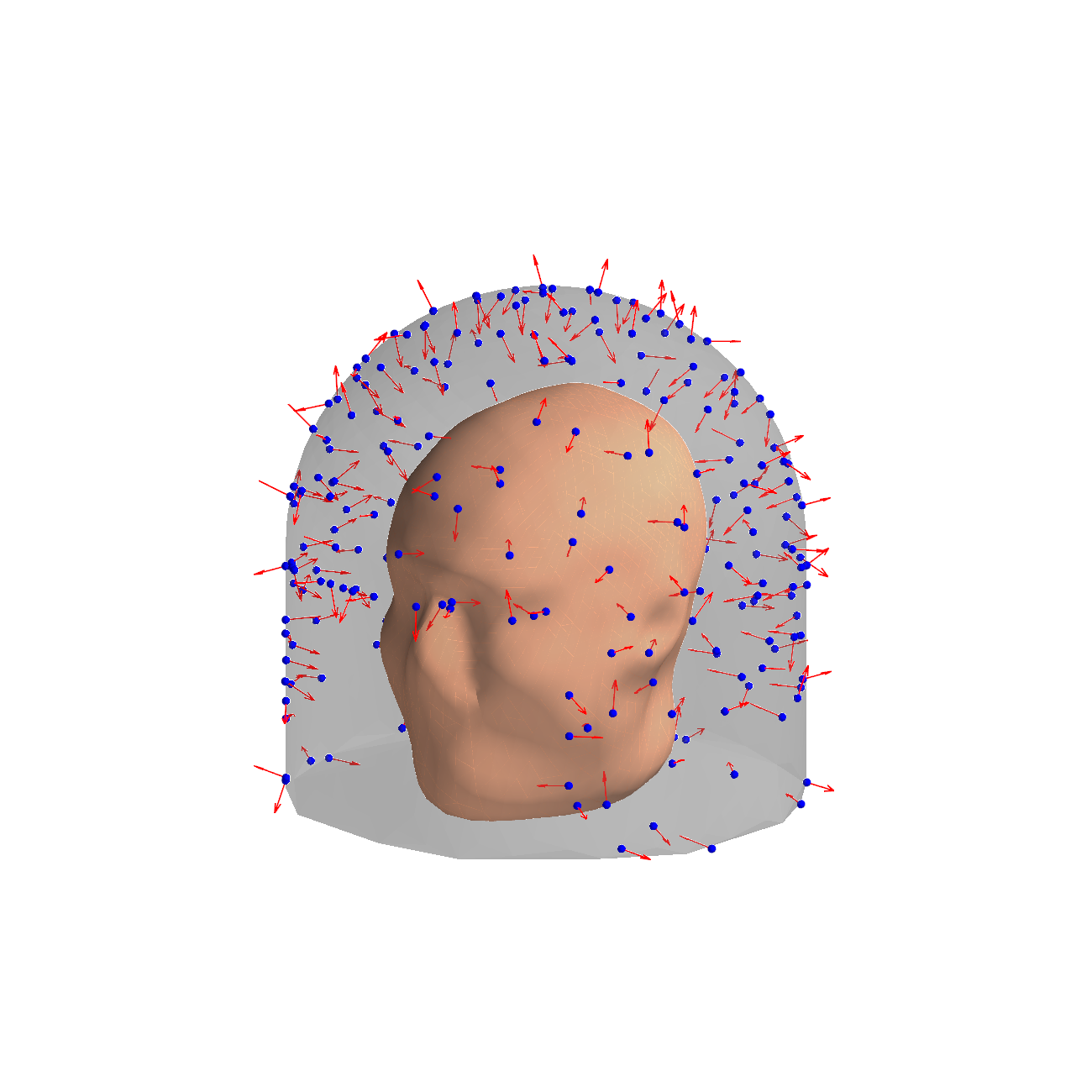}
    \end{tabular}
    \caption{Progression of the sensor arrangements during the optimization for a 2D sampling volume.}
    \label{fig:geom_opt_thin}
\end{figure}

In Figures \ref{fig:noise_opt_thin} and  \ref{fig:inf_opt_thin}, we show the results
for a 2D sampling volume, similarly to those presented in Figures \ref{fig:noise_opt}
and \ref{fig:inf_opt} for a 3D volume. Figure~\ref{fig:geom_opt_thin} depicts the
behavior of the sensor array's geometry during the optimization procedure. In this case,
noise amplification factor improves as the algorithm progresses. Similarly, channel
information capacity generally improves as a function of iteration. However, there is
a steep drop in the channel information capacity after the first iteration. The algorithm
starts with an initial configuration where the sensors are distributed uniformly on the
surface and pointing radially (Fig.~\ref{fig:geom_opt_thin}). After the first iteration,
the algorithm deviates from this configuration and the channel information capacity
decreases. However, as the algorithm progresses the channel information capacity eventually
reaches the initial level while the noise amplification factor shows an improvement
of approximately two orders of magnitude. Looking at Fig. \ref{fig:geom_opt_thin}, we
observe significant changes in the sensor orientations, which correspond to an improved
noise amplification factor and increased channel information capacity. 
%but for a 2D rather than a 3D sampling volume.

%Figure~\ref{fig:geom_opt_thin} depicts the behavior of the sensor array's
%noise amplification factor during a 2D sampling volume-based optimization procedure. Again, we
%observe significant changes in the sensor orientations, which correspond to an improved noise
%amplification factor and increased channel information capacity. 

%% file: discussion.tex
\section{Discussion}
%Discus why we are not loosing any generality by ignoring gradiometers

\subsection{General remarks}

In this paper, we investigated the question of how to optimally measure the magnetic fields in MEG with a limited number of sensors while the assumptions about the underlying neural current distribution are minimal. As a general model for a discretized, multi-channel, magnetic field measurement we use the VSH expansion. This provides us with a a tool for interpolating the magnetic field within the sampling volume. The problem of interpolating magnetic field over a curl-free domain has been studied before.
\cite{Solin2018} suggest a method for interpolating magnetic field in an arbitrarily shaped curl-free region that relies on a scalar potential function in a way that is similar to what we do. However, operating over an arbitrarily-shaped domain, the method cannot utilize the approximate spherical symmetry of the measurement geometry that we have in the MEG case.

The VSH model can be thought of as a weakly informative prior to optimize the sensor positions and orientations. The VSH prior has three parameters: the origin and the cutoff values for the inner and outer VSH expansions. The origin represents the assumption about the sensor-to-source distance while the cutoff values of the inner and outer VSH expansions constitute an assumption that the magnetic field is bandlimited in the VSH (spatial-frequency) domain. As the VSH prior is general, i.e., it is not specific to any particular subject head geometry it can be used to construct a general optimized sensor array.

One unique feature of the VSH decomposition is that it allows us to separate the field to components representing the neural signals of interest and external interference components. By using both components to construct the field model (or prior) for sensor optimization, the sensor array simultaneously samples the neural signals and the interference allowing to separate them.

Compared to previous studies on MEG array optimization using fixed sensor orientations \parencite{beltrachini2021optimal,iivanainen2021spatial}, the VSH formalism allows us to also optimize the sensor orientations. The obtained results suggest that when the sensors are measuring a single field component the optimal sensor orientation is not always radial as would be suggested when directly comparing radial component to the two tangential components (e.g., \cite{iivanainen2017measuring}). The deviation from radial orientation can be mostly explained by two different factors. First, the orientation of the spatial covariance function of the bandlimited VSH model is not radial everywhere in the sampling volume. Second, the introduction of external VSH components to the model necessitates sampling of the tangential components to allow better separation of the inner and external components.

Recently, optically pumped magnetometers (OPMs) measuring two \parencite{colombo2016four,borna201720} or all three components \parencite{brookes2021theoretical,boto2022triaxial} of the magnetic field have been developed. We did not optimize arrays comprising of triaxial or dual-axis OPMs, but we note that the methodology presented in the paper can be used to optimize such arrays.

%Introduce the idea of the virtual sensor.

%\subsection{Similarity measure for magnetic fields}
%\label{magnetic_field_similarity}
A central point of our approach to defining the MEG sensor array's figure-of-merit
is separating the question \enquote{What can we say about intracranial currents from
extracranial magnetic field measurements?} from the question \enquote{How can we
measure extracranial magnetic fields as accurately as possible?}. However, it is not
clear how these questions are related. The sensor array may sample a high percentage
of the field energy ($\sim$99$\%$, for example) giving a highly accurate
reconstruction of the magnetic field, but the source estimation might benefit
from additional sensors.

%\todo{Discuss the difference, in particular, the question of VSH normalization}

\subsection{Interpretation and significance of the obtained results}

As shown in Fig.~\ref{fig:reg_nsens}, as we interpolate the magnetic field based on the
spatially discretized measurement, the noise amplification factor decreases as
the number of sensors increases. This is an intuitively obvious result, but
Fig.~\ref{fig:reg_R} also indicates that decreasing the physical dimensions of
the actual array results in a decreasing noise amplification factor. This can
be understood by an increased density of spatial sampling as the sensors will
be distributed across a smaller surface area. 

In order to validate our approach against other metrics, we chose to compare
the progression of the noise amplification factor to the channel information
capacity, which is a commonly used quantity in the evaluation of MEG sensor
arrays. We found that  decreasing noise amplification factor during the progress
of sensor array optimization was consistent with increasing total information,
as shown in Figs \ref{fig:noise_opt} and \ref{fig:inf_opt}. As a result of the optimization
procedure, the sensors are distributed across the inner surface of the sampling
volume with widely different orientations, see Fig.~\ref{fig:geom_opt_thick}.
Similar results, with respect to the connection between noise amplification and
channel information capacity as well as the sensor orientations, were obtained
in the 2D case, as indicated in figures \ref{fig:noise_opt_thin}, \ref{fig:inf_opt_thin} and
\ref{fig:geom_opt_thin}.

It is intuitively desirable to place the sensors as close as possible to the
head with sensor normal pointing symmetrically, e.g., in the radial direction.
However, for the purpose of distinction between the internal and external magnetic
fields, it is beneficial to break the spherical measurement symmetry as much as
possible, as suggested already by \cite{nurminen2013improving}. This can be
achieved by having the sensors be close to the head while the sensor orientations
become widespread and randomly distributed. At the end of the optimization procedure
leading to these random orientations, the corresponding channel capacity returns
to the initial level as well. 

%\todo{Write the synthesis of figures 3-7 here...}

%\subsection{Correlations in the noise from "virtual" sensors}
%\todo{should this go into the limitations?}

\subsection{Other Remarks}
Note that in the process of optimization noise magnification factor drops below 1.
This means that with our sensor configuration we can estimate magnetic field
\emph{everywhere, including the sensor locations}, better than what we get by
directly measuring it with a single sensor.

\subsection{Limitations}
Our results rely heavily on the assumption that the magnetic fields within the
sampling volume can be accurately modeled with a truncated VSH expansion (see the supplementary material). This
assumption has some potential problems:
\begin{enumerate}
	\item In real MEG measurements the assumptions of the VSH expansion about the
	current geometry (three concentrical compartments) do not hold because the
	middle compartment includes a part of the participants body (neck, etc.) and
	thus cannot be guaranteed to be current-free. Moreover, for on-scalp sensor arrays, a single sphere separating the sensor array and the head cannot be found.
	\item Truncating VSH expansion naturally introduces truncation error. The
	truncation error decreases when we increase the cutoff orders for the internal
	and external parts of the expansion ($L_\alpha$ and	$L_\beta$ accordingly).
	It is not clear which cutoff values are sufficient; they depend on the SNR of the measurement.
	\item The residual VSH components of the field outside the truncated VSH expansion will alias if they are above the noise level and if the sensor array does not provide sufficient oversampling of a given truncation.
\end{enumerate}
Moreover, strictly speaking, the interpolation noise computation only accurately models
the noise for a \emph{single} "virtual" sensor. If we use it to model noise for
a virtual sensor array of multiple sensors, the noise modelling for each sensor
will be accurate, but the noise in the virtual array will be correlated across
sensors, thus the array performance will not be the same as that of a real
physical array with equivalent sensor noise.

%% file: acknowledgements.tex
\section{Acknowledgements}
The authors wish to acknowledge CSC -- IT Center for Science, Finland, for
computational resources.
 S. Taulu's work is funded in part by the Bezos Family Foundation and the R. B. and Ruth H. Dunn Charitable Foundation. This work was also supported, in part, through the Department of Physics and College of Arts and Sciences at the University of Washington.

Sandia National Laboratories is a multimission laboratory managed and operated
by National Technology \& Engineering Solutions of Sandia, LLC, a wholly owned
subsidiary of Honeywell International Inc., for the U.S. Department of Energy
National Nuclear Security Administration under contract DENA0003525. This paper
describes objective technical results and analysis. Any subjective views or
opinions that might be expressed in the paper do not necessarily represent the
views of the U.S. Department of Energy and the United States Government. The content is solely the responsibility of
the authors 